\documentclass[twocolumn,floatfix]{aastex62}

\usepackage{savesym}
\savesymbol{tablenum}
\usepackage{siunitx}
\usepackage{bm}
\usepackage{mathtools}

\graphicspath{{./}{figures/}}

\shorttitle{sBH Migration in AGN Disks}
\shortauthors{Secunda et al.}

\begin{document}

\title{Orbital Migration of Interacting Stellar Mass Black Holes in Disks Around Supermassive Black Holes}

\correspondingauthor{Amy Secunda}
\email{asecunda@princeton.edu}

\author{Amy Secunda}
\affil{Department of Astrophysics, American Museum of Natural History,
  Central Park West at 79th Street, New York, NY 10024, USA}

\author{Jillian Bellovary}
\affiliation{Department of Astrophysics, American Museum of Natural
  History, Central Park West at 79th Street, New York, NY 10024, USA}
\affiliation{Department of Physics, Queensborough Community College, Bayside, NY 11364}

\author{Mordecai-Mark Mac Low}
\affiliation{Department of Astrophysics, American Museum of Natural
  History, Central Park West at 79th Street, New York, NY 10024, USA}
\affiliation{Center for Computational Astrophysics, Flatiron
  Institute, 162 Fifth Avenue, New York, NY 10010}

\author{K.E. Saavik Ford}
\affiliation{Department of Astrophysics, American Museum of Natural
  History, Central Park West at 79th Street, New York, NY 10024, USA}
\affiliation{Department of Science, Borough of Manhattan Community
  College, City University of New York, New York, NY 10007}
\affiliation{Physics Program, The Graduate Center, CUNY, New York, NY 10016}

\author{Barry McKernan}
\affiliation{Department of Astrophysics, American Museum of Natural
  History, Central Park West at 79th Street, New York, NY 10024, USA}
\affiliation{Department of Science, Borough of Manhattan Community
  College, City University of New York, New York, NY 10007}
\affiliation{Physics Program, The Graduate Center, CUNY, New York, NY 10016}

\author{Nathan W. C. Leigh}
\affiliation{Departamento de Astronom\'ia, Facultad de Ciencias F\'isicas y Matem\'aticas,Universidad de Concepci\'on, Concepci\'on, Chile}
\affiliation{Department of Physics and Astronomy, Stony Brook University, Stony Brook, NY 11794-3800, USA}
\affiliation{Department of Astrophysics, American Museum of Natural
  History, Central Park West at 79th Street, New York, NY 10024, USA}

\author{Wladimir Lyra}
\affiliation{Department of Physics and Astronomy, California State University Northridge, 18111 Nordhoff Street, Northridge, CA 91330, USA}
\affiliation{Jet Propulsion Laboratory, California Institute of Technology, 4800 Oak Grove Drive, Pasadena, CA 91109, USA}

\author{Zsolt S\'andor}
\affiliation{Department of Astronomy, E\"otv\"os Lor\'and University, P\'azm\'any P\'eter s\'et\'any 1/A, H-1117 Budapest, Hungary}
\affiliation{Konkoly Observatory, Hungarian Academy of Sciences, Konkoly-Thege Mikl\'os \'ut 15-17, H-1121 Budapest, Hungary}

\begin{abstract}
The merger rate of stellar-mass black hole binaries (sBHBs) inferred
by the Advanced Laser Interferometer Gravitational-Wave Observatory
(LIGO) suggests the need for an efficient source of sBHB
formation. Active galactic nucleus (AGN) disks are a promising
location for the formation of these sBHBs, as well as binaries of
other compact objects, because of powerful torques exerted by the gas
disk. These gas torques cause orbiting compact objects to migrate
towards regions in the disk where inward and outward torques cancel,
known as migration traps. We simulate the migration of stellar mass
black holes in an example of a model AGN disk, using an augmented N-body code that includes analytic approximations to migration torques, stochastic gravitational forces exerted by turbulent density fluctuations in the disk, and inclination and eccentricity dampening
produced by passages through the gas disk, in addition to the standard
gravitational forces between objects. We find that sBHBs form rapidly
in our model disk as stellar-mass black holes migrate towards the migration trap. These sBHBs are likely to subsequently merge on short time-scales.  The process
continues, leading to the build-up of a population of over-massive
stellar-mass black holes. The formation of sBHBs 
in AGN disks could contribute significantly to the sBHB merger rate inferred by LIGO.

\end{abstract}


\received{2018 July 8}
\revised{2019 May 7}
\accepted{2019 May 8}

\section{Introduction}
\label{sec:intro}

The Advanced Laser Interferometer Gravitational-Wave Observatory
(LIGO) has detected the merger of stellar mass black holes (sBHs) more massive than
those previously inferred from electromagnetic observations in our own Galaxy. Additionally, while isolated binary evolution could potentially account for the high sBH merger rate inferred from LIGO detections, 52.9$^{+55.6}_{-27.0}$~Gpc$^{-3}$yr$^{-1}$ \citep{Belczynski,LIGO}, an additional mechanism of sBH mergers in the Local Universe would ease several of the assumptions necessary in these models.

It has been suggested that over-massive sBHs are most likely to form
in galactic nuclear star clusters
\citep{Hopman:2006aa,OLeary,Antonini_rasio,rodriguez}. The gas disks
in active galactic nuclei (AGN) are particularly promising locations
for the formation and merger of over-massive sBHs. As
\cite{mckernan14, mckernan18} point out, these gas disks will act to
decrease the inclination of intersecting orbiters and harden existing
binaries, already making them interesting possible locations for LIGO
detections of merging sBHs. The recent discovery of a possible black hole (BH) cusp in the core of our own Galaxy \citep{bahcall_wolf,Hailey_2018} lends further weight to this possibility.

Orbiters in a gas disk exchange angular momentum with the surrounding gas, leading to a change in semi-major axis known as migration. Migration of objects embedded within the disk provides opportunities for sBHs to form binaries if they encounter each other at small relative velocities; in particular at far smaller relative velocities than in gas-free star clusters \citep{mckernan2012,mckernan18,leigh}. If a gas disk is locally isothermal, the gas torques cause all isolated orbiters to migrate inward \citep{goldreich_tremaine,ward,tanaka}. However in the more realistic case of a disk with an adiabatic midplane, for some values of the radial density and temperature gradients
the torque from the disk can also lead to outward migration \citep{paardekooper_mellema}. 

\cite{paardekooper2010} used analytic arguments and numerical
simulations to model the sign and strength of migration, and found
that there are regions of gas disks where outward and inward torques
cancel out; leading to a region of zero net torque where migration
halts. \cite{lyra} showed that such regions of zero net torque, or
migration traps, are predicted by standard models of protoplanetary
disks, and \cite{horn} showed that the migration of protoplanets
towards these migration traps can lead to the rapid collisional
build-up of giant planet cores. 

While \citet{paardekooper2010} considered only fully
  unsaturated torques (where the angular momentum of the
  corotational region is continuously replenished by viscous mixing,
  thus continuously driving migration), updated work showed migration
  rates including saturation \citet{paardekooper2011}. The basic
  change due to saturation is twofold. First, only larger orbiters
  with mass ratio $q \gtrsim 10^{-5}$ 
  will experience sustained outward migration. For lower masses, the width of the
  horseshoe region is small enough that diffusion saturates the
  torques; rapid inward migration occurs for planets outside of a
  narrow range in mass. Second, inclusion of saturation introduces a
  mass-dependency to both the location and existence of convergence
  zones \citep{hellary_nelson, coleman, dittkrist}.  The theory of
  planet migration continues to be refined, with a dynamic 
corotation torque found \citep{paardekooper2014,pierens} dependent on the migration rate and viscosity, stemming from an asymmetry in the coorbital region as the planet moves. This torque
can stall inward and boost outward migration, taking planets away from
the convergence zone and essentially enlarging the region of outward
migration. Its action requires a Shakura-Sunyaev (\citeyear{shakura_sunyaev})
   viscosity parameter $\alpha \lesssim 10^{-2}$
for orbiters of mass ratio $\approx 10^{-5}$ (see Appendix
\ref{A}). These torques were included in N-body calculations by
\cite{sasaki_ebisuzaki}, who found
that these torques helped form cores of giant planets. Finally, a heating
torque was found by \cite{b-llambay}, resulting from the protoplanet’s
accretional luminosity, and found to counteract inward migration. The
theory of this heating torque has been further developed by
\cite{masset} and \cite{eklund_masset}, showing that it
can lead to significant eccentricity and inclination pumping. A torque formula for inclusion in evolutionary simulations has been extracted by \cite{jimenez_masset}. The state of the art in the application of these models for planet population synthesis calculations is discussed in \cite{mordasini}.

It is entirely plausible that migration models will undergo
  significant modifications in the future, driven by advances stemming
  from the unabated rate of exoplanet discoveries. Yet, some of the
  differences between AGN and protoplanetary disks cause pause, first
  and foremost the fact that the latter are relatively cold and thus
  poorly ionized, with large swaths not unstable to the MRI
  \citep{blaes,gammie,wardle,bai_stone,lesur,lyra_umurhan}. Application of planet migration theory to AGN disks should thus focus
on results for high-viscosity and turbulent gas. In this respect,
dynamical torques, requiring $\alpha \lesssim 10^{-2}$, should probably
not be too relevant (see Appendix \ref{A}). The heating torque, on the
other hand, should also exist for black hole orbiters in AGN disks:
even though they do not have a surface to heat via accretional shocks,
the accretion disks they develop are hot and luminous and should heat
up the surrounding AGN gas. We defer exploring this sBH hole feedback
effect to a future publication.

In this work, as in \citet{horn}, we prefer to work with the
  unsaturated torque because \cite{nelson}, \cite{baruteau_lin},
  \cite{uribe2011}, and \cite{baruteau2011} find that the
  co-rotational torques in turbulent disks are subject to stochastic
  turbulent fluctuations that keep the co-rotational torque
  unsaturated even in locally isothermal simulations. The result has
  been corroborated by more recent simulations
  \citep{guilet,comins,uribe}; yet, because they could not resolve the width of the corotational region for smaller objects, saturation remains a possibility if the turbulent fluctuations are strong enough to wipe out their horseshoe turns.

\citet{mckernan2012} drew on the work of \citet{lyra} and \citet{horn}
to develop a model describing a BH merger hierarchy in the AGN
disk. \cite{mckernan14} explored the consequences of this model and
predicted that LIGO should detect gravitational waves from a
previously unconsidered population of merging overweight sBH in AGN disks. \citet{bellovary} explored this analogy applying the \citet{paardekooper2010} migration torque model to two steady-state analytic supermassive black hole (SMBH) accretion disk models derived by \citet{Sirko:2003aa} and \citet{thompson}. \citet{bellovary} showed that migration traps do exist in both AGN disk models. 

Here we build on \cite{bellovary}, by using a modified version of the N-body code described by \cite{sandor} and \cite{horn} 
that implements several manifestations of the gravity of the gas disk around the SMBH in addition to the standard gravitational forces between particles.
The additional effects include migration torques, a stochastic gravitational force exerted by turbulent
density fluctuations in the disk, and inclination and eccentricity dampening produced by passages through 
the gas disk on inclined orbits\. In order to explore the dynamical behavior of multiple interacting sBHs approaching a migration trap, we take as an example the migration rates and other disk parameters derived from the analytic AGN disk model of \cite{Sirko:2003aa}. 

Embedded sBHs will migrate towards the migration traps modeled in \cite{bellovary}, and due to this migration, sBHs on prograde orbits encounter each other at low relative velocities. These encounters provide favorable conditions for fast sBHB formation and evolution, resulting in frequent mergers detectable by LIGO.
Future constraints from LIGO on this merger channel (e.g. from spins or rates) will allow us to constrain AGN disk physics better than present spectroscopic modeling efforts (see \cite{mckernan18} for a discussion of which parameters can be best constrained by LIGO).

\section{Methods}
\label{sec:methods}

In this section we describe in detail our modified N-body simulations. Our simulations neglect forces exerted by sBHs on the gas disk aside from those implicitly modeled by the migration torques, the effects of accretion onto either the central SMBH or orbiting sBHs, and general relativistic effects. We also only consider sBHs on prograde orbits and ignore sBHs on retrograde orbits around the central object. We defer detailed modeling of retrograde objects until the torques on them have been derived in work in progress.
 
\begin{figure}[htb!]
\label{fig:sirkoM}
\includegraphics[width=0.5\textwidth]{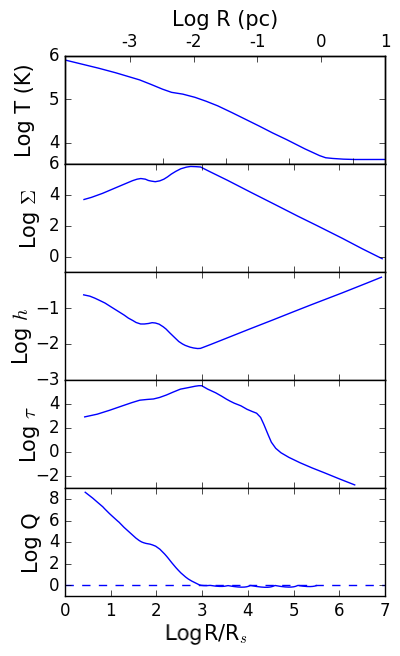}
\caption{SMBH accretion disk model used in our simulations \citep{Sirko:2003aa}. From top to bottom are plotted the midplane temperature $T$, surface density $\Sigma$ (in g~cm$^{-2}$), disk aspect ratio $h$ ($H/r$), optical depth $\tau$, and Toomre $Q$ as a function of Schwarzschild radius $R_{\rm s}$. The top axis represents the translation from Schwarzschild radius to parsecs for a $10^8$~M$_{\sun}$ SMBH.}
\end{figure}

\subsection{Disk Models}
\label{sec:sirko}

The \citet{Sirko:2003aa} model is a modification of the classic Keplerian viscous
disk model \citep{shakura_sunyaev}, with a constant high accretion rate fixed at Eddington ratio 0.5. The disk is assumed to be marginally stable to gravitational fragmentation;
however the model does not directly take into account magnetic fields or general relativistic effects. The \citet{Sirko:2003aa} model assumes some additional unspecified heating mechanism in the outer disk in order to maintain the stability of the disk and prevent fragmentation. 

\citet{Sirko:2003aa} use the opacity models from \cite{iglesias_rogers} and \cite{alexander_ferguson} for the high and low temperature regimes, respectively. The inner disk is optically thick due to a high rate of Thompson scattering from electrons produced by the ionization of hydrogen. The intermediate region of the disk has a lower electron density, and is therefore less optically thick and cooler. 

We use a SMBH mass of $M_{\star}$ = $10^8$~M$_{\sun}$. The total mass of the disk integrated out to \num{2e5}~AU is $\num{3.7e7}$~M$_{\sun}$. The midplane temperature, surface density, scale height, optical depth, and Toomre Q as a function of radius in this model are plotted in Figure \ref{fig:sirkoM}. 

\subsection{Torque Model}
\label{sec:migration}
We model the disk torque on the sBHs using the analytical prescription of \cite{paardekooper2010} which incorporates the effects of non-isothermal co-rotation torques. For the azimuthally isothermal case the normalized torque is
\begin{equation}
    \Gamma_{\rm iso}/\Gamma_{0}= -0.85 - \alpha - 0.9\beta,
	\label{eq:iso torque}
\end{equation}
while for the purely adiabatic case the normalized torque is
\begin{equation}
    \gamma\Gamma_{\rm ad}/\Gamma_{\rm 0}= -0.85 - \alpha - 1.7\beta +7.9\xi/\gamma.
	\label{eq:ad torque}
\end{equation}
The adiabatic index $\gamma = 5/3$, and the variables $\alpha$,
$\beta$ and $\xi$ represent the negative local gradients of density,
temperature and entropy, respectively, and are defined as
\begin{equation}
    \alpha = -\frac{\partial \ln\Sigma}{\partial \ln r};\,\, \beta = -\frac{\partial \ln T}{\partial \ln r};\,\, \xi = \beta - (\gamma - 1)\alpha.
	\label{eq:gradients}
\end{equation}
The torques are normalized by
\begin{equation}
	\Gamma_0 = (q/h)^2 \Sigma r^4 \Omega^2,
    \label{eq:normT}
\end{equation}
where $q$ is the mass ratio of the migrator to the SMBH, $h$ is the
aspect ratio of the disk and $\Omega$ is the rotational velocity. 

The effective torque is interpolated between the isothermal and
adiabatic torque models using 
	\begin{equation}
		\Gamma = \frac{\Gamma_{\rm ad}\Theta^2 + \Gamma_{\rm iso}}{(\Theta + 1)^2},
	\end{equation}
where $\Theta$ is the ratio of the radiative timescale to the dynamical timescale. \cite{lyra} show that $\Theta$ depends on the local disk properties as
\begin{equation}
	\Theta = \frac{c_{\rm v}\Sigma\Omega\tau_{\rm eff}}{12\pi\sigma T^3},
\end{equation}
where $c_{\rm v}$ is the thermodynamic constant at constant volume, $\sigma$ is the Stefan-Boltzmann constant, and the effective optical depth taken at the midplane is \citep{hubeny,kley_crida}
\begin{equation}
\label{eq:taueff}
	\tau_{\rm eff} = \frac{3\tau}{8} + \frac{\sqrt[]{3}}{4} + \frac{1}{4\tau}.
\end{equation}
The true optical depth $\tau$ is given by
\begin{equation}
\label{eq:tau}
	\tau=\frac{\kappa\Sigma}{2}
\end{equation}
where $\kappa$ is the opacity used in the \cite{Sirko:2003aa} models (see Section \ref{sec:sirko}). 

Each component of the torque depends on the local disk gradients of density, temperature and entropy. 
These torques are implemented into our N-body code as forces on the particles with vector dependence
\begin{equation}
\label{eq:fmig}
\bm{F}_{\rm mig} = \frac{\Gamma}{r}\hat{\theta}
\end{equation}

\subsection{Turbulence}
\label{sec:turbulence}
AGN disks are sufficiently ionized (certainly in the inner regions) that the magnetorotational instability (MRI) will drive
turbulence. We use a model for turbulence developed by \cite{Laughlin_1994} and further modified by \cite{ogihara} that gives
the gravitational forces exerted by turbulent density fluctuations 
as
	\begin{equation}
	\label{eq:fturb}
	\bm{F}_{\rm turb} = -C\nabla \Phi,
	\end{equation}
where \textit{C} is a scaling factor relating the fraction of the force exerted on the gas by the potential $\Phi$ to the force that is exerted by the gas on a migrator embedded in the disk. This fraction is given as 
	\begin{equation}
	\label{eq:C}
	C = \frac{64\Sigma r^2}{\pi^2 M_{\star}}.
	\end{equation}

The turbulent potential, $\Phi$, is taken to be the sum of $n = 200$ independent, scaled oscillation modes
	\begin{equation}
	\label{eq:tpot}
	\Phi_{\rm c,m} = \psi r^2 \Omega^2 \Lambda_{\rm c,m},
	\end{equation}
where $\psi$ is a dimensionless measure of the strength of the
turbulent force in comparison to the migration forces (see Section
\ref{sec:migration}).  It is related to the \cite{shakura_sunyaev} viscosity parameter $\alpha$ by \citet{baruteau_lin} as
	\begin{equation}
	\label{eq:gamma}
	\psi \simeq \num{8.5e-2} h  \alpha^{1/2}
	\end{equation}
where $h$ is the aspect ratio of the disk and comes from the mode lifetime being set by the speed of sound. In our model $h$ is not constant, but to fix the scaling
in Equation~(\ref{eq:gamma}) we set $h=0.05$. MHD simulations of accretion disks suggest typical values for $\alpha$ of $10^{-3}$--0.1 \citep{davis}. A value of $\alpha = 0.01$ gives us $\psi = \num{4.25e-4}$.

In Equation (\ref{eq:tpot}), $\Lambda_{\rm c,m}$ is one oscillation mode defined as
	\begin{equation}
	\label{eq:osc_mode}
	\Lambda_{\rm c,m} = \xi e^{-(r-r_{\rm c})^2/\sigma^2} \cos(m\theta - \phi_{\rm c} - \Omega_{\rm c} \tilde{t}) \sin\left(\pi \frac{\tilde{t}}{\Delta t}\right).
	\end{equation}
Each oscillation mode is defined by $m$, an azimuthal wavenumber
chosen from a log normal distribution between 1 and 64, and $c$ denotes the
initial center of the perturbation. The position $c$ is given in cylindrical coordinates $r_{\rm c}$ and $\phi_{\rm c}$ selected from uniform distributions from the inner boundary to the outer boundary of the disk and from 0 to $2\pi$, respectively. The $z$ coordinate is assumed to be small enough to have a negligible effect. $\Omega_{\rm c}$ is the Keplerian angular velocity at $r_{\rm c}$. 

The mode evolves as a function of $\tilde{t} = t_{\rm 0} + t$, where $t_{\rm 0}$ is the time when the mode comes into existence. The lifetime of the perturbation is
	\begin{equation}
	\label{eq:deltaT}
	\Delta t = \frac{2\pi r_{\rm c}}{mc_{\rm s}},
	\end{equation}
which represents the sound-crossing time for each mode. The radial scale of the perturbation is chosen from a Gaussian distribution and scales as $\sigma = \pi r_{\rm c}/4m$. 

At the beginning of the simulation there are $n = 200$ modes. When one mode expires another mode is created so that there are always 200 modes. \cite{ogihara} showed that all modes $m > 6$ can be left out of the summation to determine the total potential $\Phi$. We use this simplification in our model and only include $\Phi$ perturbations where $m < 7$. Equation~(\ref{eq:fturb}) is used in our model to calculate the turbulent force on a given migrator at position $(r, \theta)$ as a function of the local speed of sound, Keplerian angular velocity, surface density of the gas, and time.

We note that when the net vertical magnetic flux of the disk is not sufficiently large, spiral acoustic waves or even radiation stresses dominate angular momentum transport and accretion power instead of MRI turbulence \citep{jiang_stone}. While the perturbations generated through these mechanisms will not be identical to those produced by MRI turbulence, as modeled above, we anticipate they will have qualitatively the same effect on our simulations (see Section \ref{sec:results}).

\subsection{Eccentricity and Inclination Dampening}
\label{sec:dampening}
\cite{tanaka_ward} have shown that the gas disk exerts a force on migrators that acts to dampen their orbital eccentricity, $e$, and inclination, $i$, leading to the co-planar circularization of orbiters. They give the timescale
\begin{equation}
\label{eq:tdamp}
	t_{\rm damp} = \frac{M_{\rm \star}^2 h^4}{m \Sigma a^2 \Omega},
\end{equation}
where $m$ is the mass of the migrator and $a$ is the semimajor axis of the migrator. We follow the timescales given in \cite{cresswell_nelson} for eccentricity and inclination, respectively:
	\begin{equation}
	\label{eq:edamp}
    t_{\rm e} = \frac{t_{\rm damp}}{0.780}(1 - 0.14\epsilon^2 + 0.06\epsilon^3 + 0.18\epsilon l^2)
    \end{equation}
    \begin{equation}
    \label{eq:idamp}
    t_{\rm i} = \frac{t_{\rm damp}}{0.544}(1 - 0.30l^2 + 0.24l^3 + 0.14l\epsilon^2)
	\end{equation}
where $\epsilon = e/h$ and $l=i/h$. 

The resulting forces acting on these timescales as a function of position and velocity of an orbiting body are 
\begin{equation}
\label{eq:fdampr}
\boldsymbol{F}_{\rm damp,r} = -2 \frac{(\boldsymbol{v} \cdot \boldsymbol{r})\boldsymbol{r}}{r^2 t_{\rm e}}m\boldsymbol{\hat{r}}
\end{equation}
\begin{equation}
\label{eq:fdampz}
\boldsymbol{F}_{\rm damp,z} = - \frac{v_{\rm z}}{t_{\rm i}}m\boldsymbol{\hat{z}},
\end{equation}
where $\boldsymbol{\hat{r}}$ and $\boldsymbol{\hat{z}}$ are unit vectors in the $r$ and $z$ directions, respectively.

\subsection{N-Body Code}
\label{sec:nbody}

We use the Bulirsch-Stoer N-body code described by \cite{sandor} that was modified by \cite{horn} to include the additional forces outlined above in Sections \ref{sec:migration}, \ref{sec:turbulence}, and \ref{sec:dampening}. The total force acting on each sBH in our simulation is
\begin{equation}
\label{eq:ftotal}
\boldsymbol{F}_{\rm total} = \boldsymbol{F}_{\rm nbody} + \boldsymbol{F}_{\rm mig} + \boldsymbol{F}_{\rm damp} + \boldsymbol{F}_{\rm turb}.
\end{equation}

The forces acting from the gas disk, $\boldsymbol{F}_{\rm mig}$, $\boldsymbol{F}_{\rm damp}$ and $\boldsymbol{F}_{\rm turb}$, are calculated at the beginning of each Bulirsch-Stoer timestep and not recalculated during the modified midpoint method used to calculate $\boldsymbol{F}_{\rm nbody}$. However, the Bulirsch-Stoer timestep is a small fraction of the dynamical timescales of the sBHs and is reduced during close encounters. Therefore holding these forces from the gas disk constant throughout each Bulirsch-Stoer timestep does not have a significant effect on the simulations.

Our simulations consider two sBHs to have formed a new sBHB once two conditions have been met. First, they must approach each other within a mutual Hill radius,
 	\begin{equation}
	\label{eq:rhill}
	R_{\rm mH} = \left(\frac{m_{\rm i} + m_{\rm j}}{3M_{\rm \star}}\right)^{1/3} \left(\frac{r_{\rm i} + r_{\rm j}}{2}\right),
	\end{equation}
where $m_{\rm i}$ and $m_{\rm j}$ represent the masses of the two sBHs and $r_{ \rm i}$ and $r_{\rm j}$ represent their distances from the SMBH.  Second, the relative kinetic energy of the binary,
	\begin{equation}
	\label{eq:KE}
    K_{\rm rel}=\frac{1}{2}\mu v_{\rm rel}^2,
	\end{equation}
where $\mu$ is the reduced mass of the binary, and $v_{\rm rel}$ is the relative velocity between the two sBHs, must be less than the binding energy,
	\begin{equation}
	U=\frac{G m_{\rm i} m_{\rm j}}{2R_{\rm mH}}.
	\end{equation}

Due to the complex and poorly understood interactions between sBHBs and the gas disk within the Hill sphere, for simplicity once a gravitationally bound sBHB forms, our model assumes that it is merged. Indeed it is likely given the conditions of our simulations that all sBHBs will merge within approximately 10--500 yr \citep{baruteau_lin}, which is a short timescale compared to any dynamical timescales. However, escapes from within a mutual Hill sphere are of course possible. We discuss the merging of sBHBs in our simulations further in Section \ref{sec:discuss}. 

\section{Initial Stellar Mass BH Populations}
\label{sec:initial}
In this section we describe the two models for the initial sBH populations
used in our simulations, which are outlined in Table
\ref{table:runs}. We choose the number of sBHs in each model based on the lower limit of about $10^3$ sBHs within 0.1 pc of a SMBH estimated 
by \citet{antonini} based on the distribution of S-Star orbits around Sgr A$^{\star}$. This estimate is consistent with the population
of O($10^{4}$) sBHs within 1~pc of Sgr A$^{\ast}$ inferred by
\cite{Hailey_2018}. Assuming sBHs are uniformly distributed throughout
the disk, we estimate that around 1\% of sBHs in an AGN disk will be
within the inner 1000 AU ($\approx$0.005 pc). Both of our models therefore include ten sBHs within roughly 1000 AU.

The gravitational wave decay lifetime of a sBH a few hundred AU from a SMBH in a gas-free nucleus is \citep{peters},
\begin{equation}
\label{eq:GWdecay}
T(a_{\rm 0},e_{\rm 0}) \approx \frac{768}{425}\frac{(1-e_{\rm 0}^2)^{7/2}a_{\rm 0}^4}{4\beta},
\end{equation}

where $\beta$ is,

\begin{equation}
\label{eq:beta}
\beta = \frac{64}{5}\frac{G^3m_{\rm 1}m_{\rm 2}(m_{\rm 1}+m_{\rm 2})}{c^5}.
\end{equation}

Using $m_{\rm 1}$ = $10^8$ $M_{\rm \sun}$, $m_{\rm 2}$ = 30 $M_{\rm \sun}$, $e_{\rm 0}$ = 0.05, and $a_{\rm 0}$ = 650 AU as fiducial values that are used in our runs (see below), gives a decay time of approximately \num{3.72e11} yr. Since this value is several orders of magnitude longer than the run time of our simulations, our models do not include the gravitational wave decay of the orbits of the sBHs around the SMBH.

\begin{table}[htb!]
	\centering
	\caption{Models.
Column 1: Name of run; Column 2: initial masses (or range of masses) of bodies in $M_{\rm \sun}$; Column 3: the total combined mass of all bodies in the run in $M_{\rm \sun}$; Column 4: the time it takes for all bodies to reach the migration trap or resonant orbits in megayears; Column 5: the time for a sBHB of over 50 $M_{\rm \sun}$ to form in megayears; Column 6: the mass in $M_{\rm \sun}$ of the most massive sBH at the end of the run.}
	\label{table:runs}
	\begin{tabular}{lccccr}
		\hline
		Run & $M_{\rm sBH}$ & $m_{\rm tot}$ & $T_{\rm mig}$ & $T_{\rm form}$ & $m_{\rm max}$\\
        (1) & (2) & (3) & (4) & (5) &(6) \\
		\hline
		F1 & 10 & 100 & 0.14 & 0.129 & 70\\
		F2 & 20 & 200 & 0.025 & 0.008 & 100\\
        F3 & 30 & 300 & 0.014 & 0.002 & 240\\
		\hline
        LMA & 5--15 & 74 & 0.7 & N/A & 46\\
        LMB & 5--15 & 100 & 2.8 & 1.5 & 65\\
        HMA & 5--30 & 97 & 0.45 & 0.24 & 60\\
        HMB & 5--30 &95 & 0.8 & 0.56 & 59\\
	\end{tabular}
\end{table}

Our three fiducial models (labeled F1--F3 in Table \ref{table:runs})
contain 10 sBHs of uniform masses. This uniform mass distribution is
different for each fiducial model, and ranges from 10 $M_{\rm \sun}$
in F1 to 30 $M_{\rm \sun}$ in F3. The innermost sBH has an initial
semi-major axis of 500 AU. The semi-major axis of each successive sBH
is separated by 30 $R_{\rm mH}$ from the one before it (see Equation
\ref{eq:rhill}). These initial positions are chosen to create a
distribution of sBHs around the migration trap found at roughly 667 AU
by \cite{bellovary}. We note that this initial distribution is
  somewhat arbitrary, however, these fiducial runs are mainly used as a baseline example
to show how sBHs of different masses and initial semi-major axes that are initially not under each other's gravitational influence can migrate to form sBHBs in an AGN disk.

In our second set of models the masses of the sBHs vary in a more physically realistic manner.  We draw them
from the initial mass function for massive stars given by \cite{kroupa}, by drawing from a Pareto power law probability distribution of sBHs with a probability density 
\begin{equation}
\label{eq:imf}
p(x) = \frac{am_{\rm 0}^a}{x^{a+1}},
\end{equation}
where $a = 1.35$, $m_{\rm 0}$ is a scale factor of 5 M$_{\rm \sun}$, and $x$ is a mass that is drawn from the distribution. 

In our two lower mass runs, denoted LMA and LMB, a randomly generated
mass is rejected if it is greater than 15 $M_{\rm \sun}$, so the masses
of the sBHs range from 5--15 $M_{\rm \sun}$. In our two higher mass
runs, denoted HMA and HMB, the mass is allowed to range from 5--30
$M_{\rm \sun}$. Despite being denoted higher and lower mass runs,
the total mass of the higher mass runs does not always exceed that of the lower mass runs because of random variation. This is the case for LMB, for example, which has the highest total mass of 100
$M_{\rm \sun}$. The initial semimajor axes for the sBHs in these models
are chosen randomly from a uniform distribution ranging from 300--1000
AU. We do not use an initial-final mass relation for the sBHs \citep[i.e.][]{fryer} which would require us to make assumptions of the metallicity and supernova explosion model of our simulations. However, our distribution of initial masses for our sBHs remains similar to what they would be if such a relation had been used.

For all models, the initial eccentricities and inclinations of the
sBHs are selected randomly from a Gaussian distribution. The mean
value for the initial eccentricity is 0.05, with a standard deviation
of 0.02. Selections are made until the value is positive.
The mean value for the inclination is 0 with a standard deviation of $\ang{0.05}$, and the absolute value of the randomly selected value is used. The initial mean anomaly and pericenter values are chosen randomly from a uniform distribution ranging from 0 to $2\pi$.

For the variable mass models the distance between sBHs is calculated based on the randomly generated positional coordinates. If any two sBHs are within 10 AU of each other, a new distribution is generated until no two sBHs are within 10 AU of each other.

The masses of the sBHs remain constant over the course of the simulations, i.e. the sBHs are not accreting gas. This is a realistic simplifying assumption based on the Eddington-limited accretion rates, which would give a mass doubling time of about 40~Myr. Since our simulations are only run for 10~Myr and most of the mergers take place within the first few megayears, the additional mass due to accretion is insignificant, to both the mass of the object, and the migration rate. Gas accretion onto the sBHs could have a significant effect on the gas disks around the sBHs (i.e. feedback). However, these back reactions have not been well quantified and so we defer the study of the effects of gas accretion to future work.

These models were run for 10 Myr which is within the range of estimated lifetimes for an AGN disk \citep{haehnelt,king_nixon,schawinski}.
However, the final orbits of all sBHs in all seven models are established in less than 3 Myr, and these orbits remain stable for the remainder of the run. Over
longer periods of time we would expect more sBHs to migrate inwards
towards the SMBH from the outer disk. These sBHs may perturb the
stable resonant orbiters or sBHs in the migration trap. We defer
investigation of this evolution to future work.

\section{Results}
\label{sec:results}

\begin{figure*}
\centering
\begin{tabular}{cc}
\includegraphics[width=0.95\textwidth]{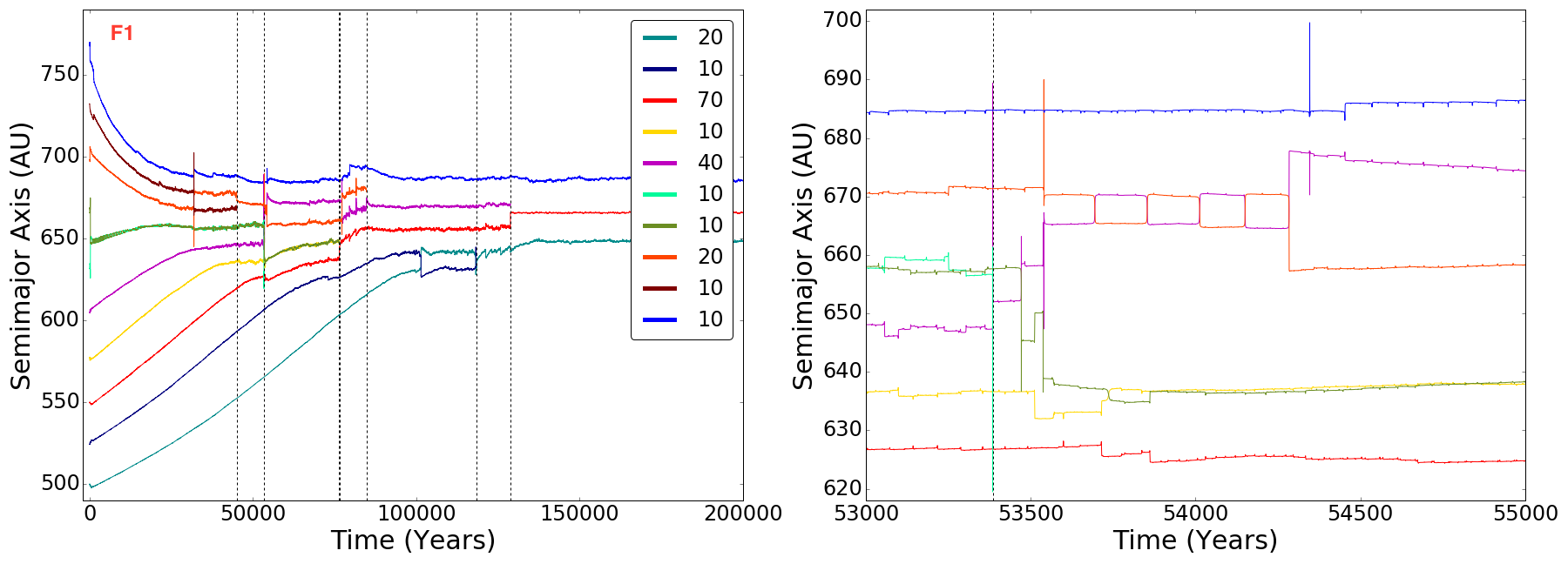} \\
\includegraphics[width=0.95\textwidth]{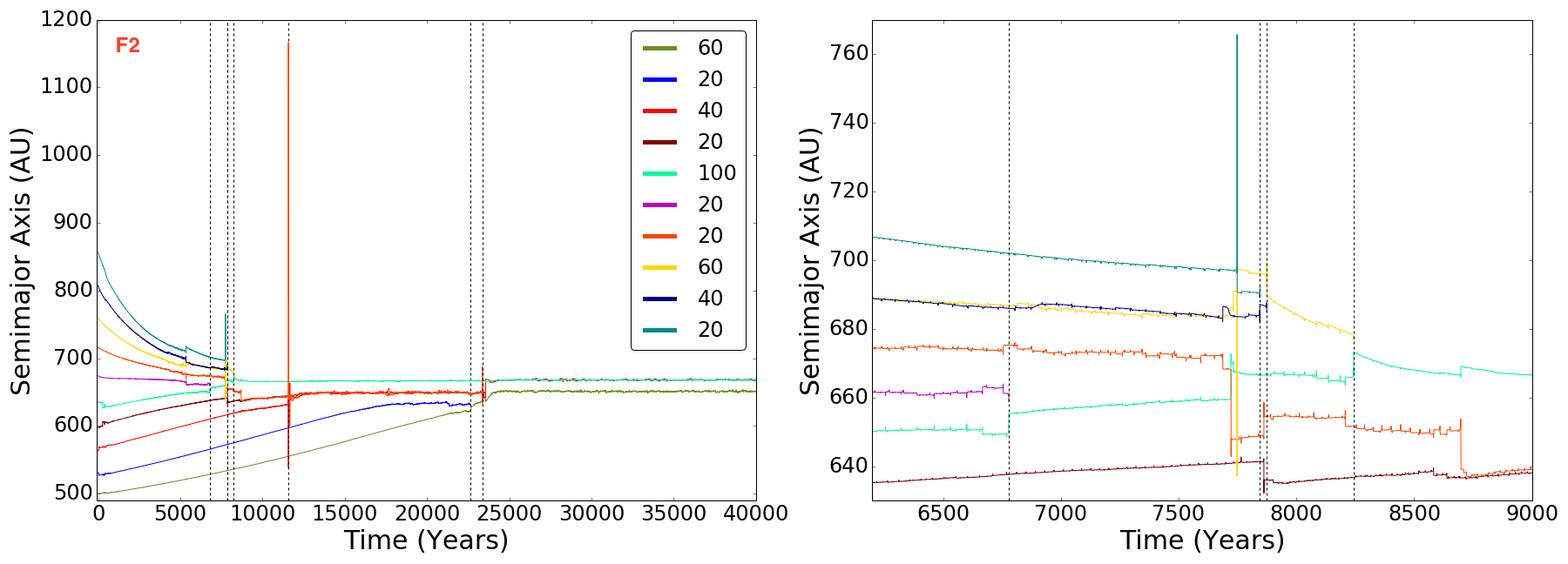} \\
\includegraphics[width=0.95\textwidth]{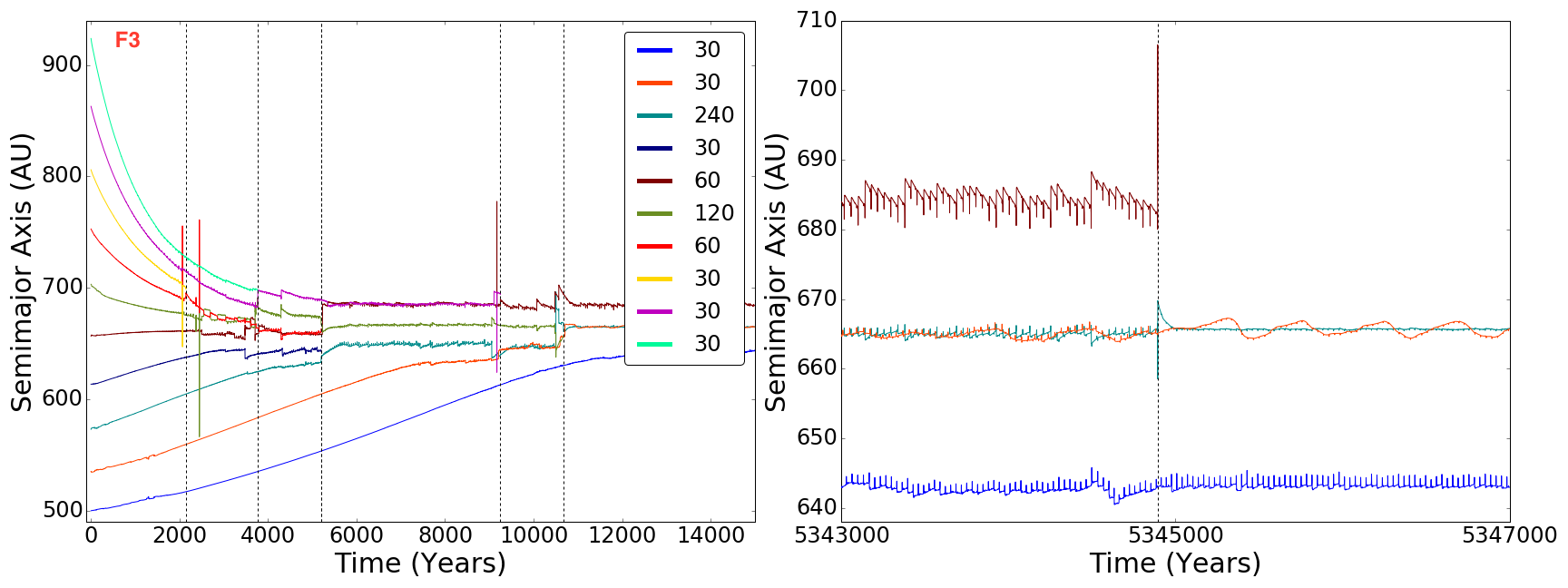}
\end{tabular}
\caption{The migration of ten sBHs for all three fiducial runs. The
  initial masses of the sBHs are 10 M$_{\rm \sun}$ (top), 20 M$_{\rm \sun}$
  (middle), and 30 M$_{\rm \sun}$ (bottom). Each colored line represents
  one sBH and is labeled by its final mass in M$_{\rm \sun}$. Each
  vertical dashed black line represents a time at which a bound binary
  forms. The figures on the left show the main period during which
  binary formation occurs. In the top two panels the sBHs remain on
  the same orbits that they are on at the end of the figures for the
  remainder of the simulations. In the bottom panel turbulence knocks
  a sBH out of resonance after roughly 5 Myr (see bottom right
  panel). The figures on the right show zoomed in views of various episodes 
of binary formation. The 100 $M_{\rm \sun}$ and 40 $M_{\rm \sun}$ sBHs in the center left panel and the 240 $M_{\rm \sun}$ and 30 $M_{\rm \sun}$ in the bottom right panel end up on the trojan orbits discussed in Section \ref{sec:results:kroupa}}
\label{fig:pathR_fid}
\end{figure*}

\begin{figure}[htb!]
\begin{tabular}{l}
\includegraphics[width=0.5\textwidth]{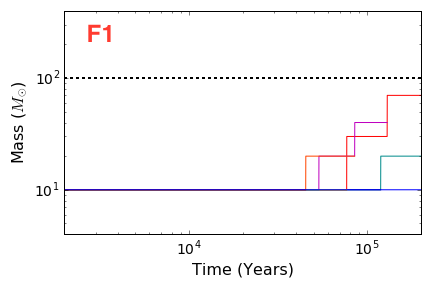} 
\\ \includegraphics[width=0.5\textwidth]{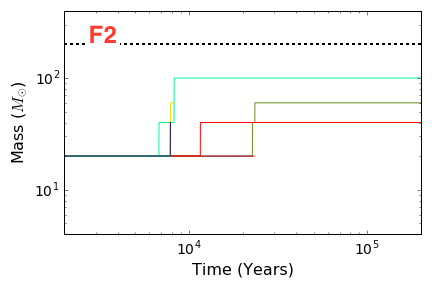}
\\ \includegraphics[width=0.5\textwidth]{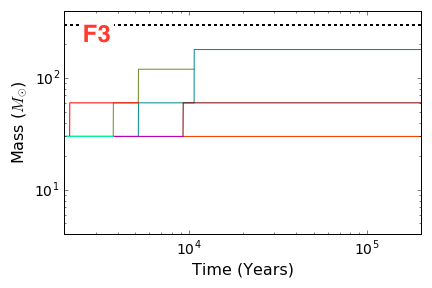}
\end{tabular}
\caption{The masses of the sBHs over time for runs F1, F2, and F3 are shown in the top, middle, and bottom figures, respectively. Each colored line represents a sBH. The dashed black line represents the total mass of all sBHs in the model.}
\label{fig:growth_fid}
\end{figure}

\begin{figure*}[ht!]
\includegraphics[width=\textwidth]{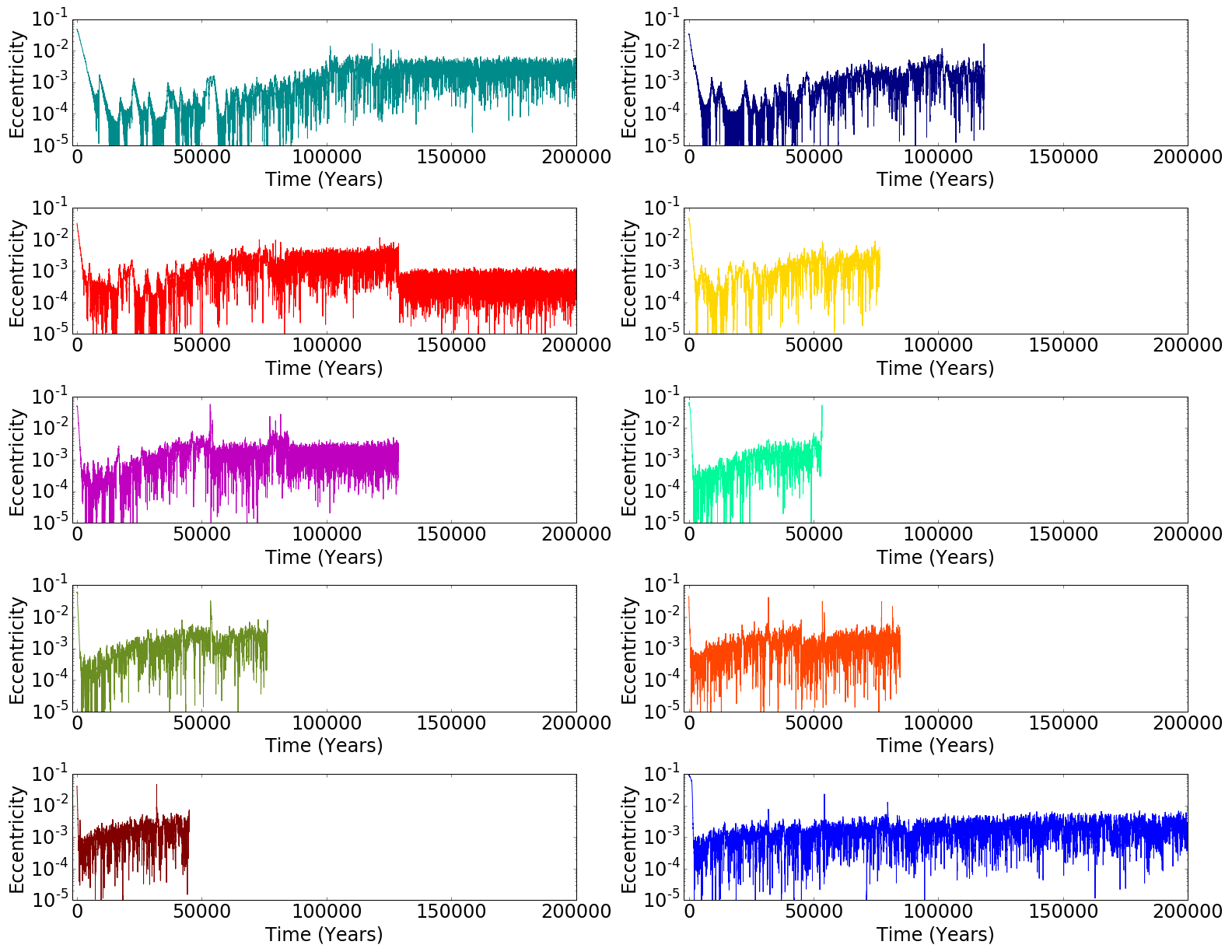}
\caption{The eccentricities over the first 200 kyr of all ten sBHs in the F1 run. Initial eccentricities are quickly dampened by the gas in less than 10 kyr, however interactions between the sBHs in close proximity drives the eccentricity of the sBHs' orbits as they are pulled towards each other. This eccentricity is then dampened, until another close passage occurs.}
\label{fig:ecc}
\end{figure*}

\subsection{Fiducial Model}
\label{sec:results:fiducial}
Figure \ref{fig:pathR_fid} shows the migration history for runs F1, F2, and F3 in the top, middle, and bottom panels, respectively (see Table \ref{table:runs}). In the top two panels of Figure \ref{fig:pathR_fid} the figures on the left show the migration history from the start of the run to shortly after the final merger. The orbits of the remaining sBHs stay the same until the end of the 10 Myr run. The figures on the right in the top two panels are zoomed in views of mergers for runs F1 and F2. 

The bottom left panel of Figure \ref{fig:pathR_fid} shows the main period of mergers for the F3 run. The orbits of the four remaining sBHs remain the same for over 5 Myr. However, a turbulent mode (see Section \ref{sec:turbulence}) opens up near the remaining orbiters at around 5.3 Myr causing the 60 $M_{\rm \sun}$ sBH to form a sBHB with the 180 $M_{\sun}$ sBH. This merger is shown in the bottom right panel of Figure \ref{fig:pathR_fid}. The turbulent mode continues to cause the semimajor axes of the orbits of the sBHs' in the migration trap to oscillate. The oscillations are more distinctive in the semimajor axis of the 30 $M_{\rm \sun}$ sBH, because it is significantly less massive than the 240 $M_{\rm \sun}$ sBH.

In these fiducial runs it is clear that more massive bodies migrate
faster towards the migration trap, as expected since the migration
torque is proportional to the square of the mass of the orbiter and so
the acceleration is linearly proportional to mass. Thus, more massive
sBHs reach the migration trap more rapidly. For example, the sBHs in
model F3 all reach the migration trap or nearby resonant orbits in roughly 14 kyr, whereas it takes the sBHs in model F1 around 140 kyr. In all cases the last sBHs to reach the migration trap region are the innermost sBHs. These innermost sBHs have the slowest migration rates because within 1000 AU of the SMBH the aspect ratio of the disk increases with proximity to the SMBH (see Figure \ref{fig:sirkoM}). The higher aspect ratio of the inner disk also means that the innermost sBHs will remain on eccentric orbits longer than sBHs since the damping force, ${F}_{\rm damp,r}$ is inversely proportional to $h^4$ (see Equations \ref{eq:tdamp} - \ref{eq:fdampr}).

Figure \ref{fig:growth_fid} shows the growth of sBHs through mergers over time. Massive bodies approaching or reaching the migration trap encounter each other at high rates. Since binaries form at greater rates as sBHs migrate towards the migration trap, the faster migration rate of the more massive bodies leads to faster sBHB formation in the more massive fiducial models. For example, F2 and F3 both have four sBHBs form within the first 10 kyr, whereas it takes nearly 50 kyr for a sBHB to form in F1.

Figure \ref{fig:ecc} shows the eccentricity of all ten sBHs over the
first 200 kyr for the F1 run. While the initial eccentricities of the sBHs' orbits are dampened by the gas within the first 10 kyr, these eccentricities can actually delay
sBHB formation at earlier times in our simulations. sBHs that are on
eccentric orbits may pass within a Hill radius of each other, but because their orbits have different pericenter phases their relative velocities are great enough that the relative kinetic energy of the two sBHs remains greater than their binding energy (see Section \ref{sec:nbody}).

Oscillations in the eccentricity of the orbits of the sBHs that occur
later in the run are due to interactions between sBHs. As the sBHs
migrate into closer proximity with each other they will be pulled
towards each other. This feature can be seen in Figure
\ref{fig:pathR_fid} as little spikes in the semimajor axes of the
orbiters. These spikes can be periodic if they occur when two orbiters with similar semimajor axes are in phase with each other. The change in semimajor axis drives the eccentricity of the sBHs. The gas disk will dampen these eccentricities, leading to a decrease in eccentricity until another close pass occurs. These interactions are what cause the oscillations in Figure \ref{fig:ecc}. The eccentricity of the sBH orbits rarely increases to more than $10^{-2}$. This feature is common in our simulations and is discussed further in Section \ref{sec:results:kroupa}. 

When sBHs pass within just a couple of Hill radii 
of each other, whether on not their relative kinetic energy is low
enough to form a sBHB, the effective 
semimajor axis of their orbits around the
SMBH often spike dramatically as their orbits are strongly perturbed from Keplerian orbits, 
as can be seen in Figure \ref{fig:pathR_fid}. However, this should be interpreted as a dramatic change in velocity rather than position.

In our runs the most massive sBH consistently ends up closest to the
migration trap. However, in some cases, such as in the F2 run, no sBH
ends up precisely in the migration trap. Instead the most massive sBH ended 
up roughly 2.5~AU away from the migration trap. At these small
distances, the migration torque is very minimal, and the dynamics due
to the high density of sBHs in the region play a larger role in
determining the orbits' positions. Less massive sBHs end up either on
Trojan or resonant orbits that exchange angular momentum with the
other sBHs. These final configurations tend to be stable on megayear time scales. However, MRI turbulence can lead to sBHB formation even after these stable orbits are established if it knocks a sBH out of resonance, as happened in the F3 run. Encounters with other objects either being ground down into the disk, or migrating inward from further out in the disk might also disturb the steady configurations over longer time scales.

\subsection{Varying Masses}
\label{sec:results:kroupa}

\begin{figure}
\includegraphics[width=0.455\textwidth]{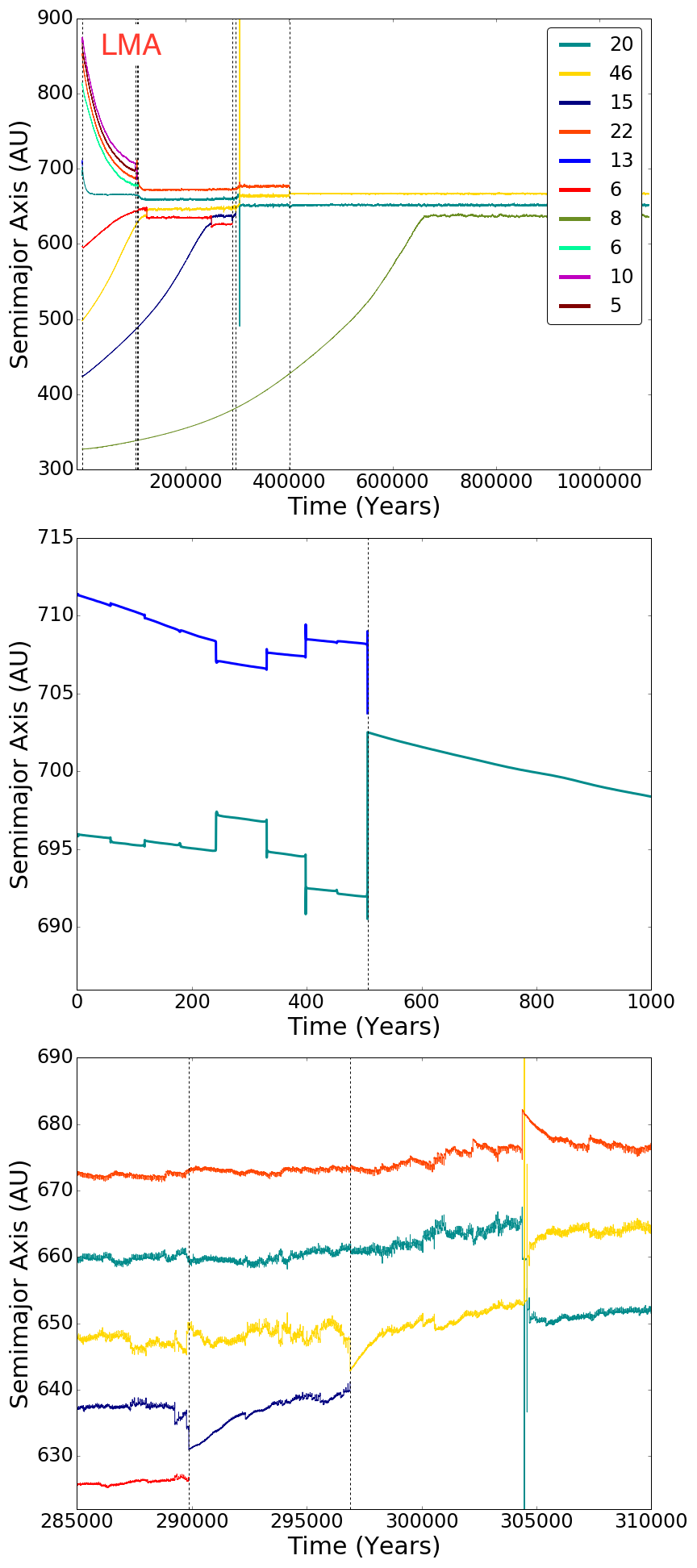}
\caption{The migration of ten sBHs of varying mass in model LMA. Each
  colored line represents one sBH and is labeled by its final mass in
  M$_{\rm \sun}$. Each vertical dashed black line represents a time at
  which a collision occurs. The top figure shows the first 1.1 Myr
  which is the period during which binary formation occurs, and all
  sBHs migrate towards the migration trap to stable orbits where they
  remain for the rest of the 10 Myr run. The middle figure is a zoomed
  in view of the first binary capture (so early that it is barely visible in the top panel) and the bottom figure is a zoomed in view of a later period.}
\label{fig:pathRLMA}
\end{figure}

\begin{figure}
\includegraphics[width=0.47\textwidth]{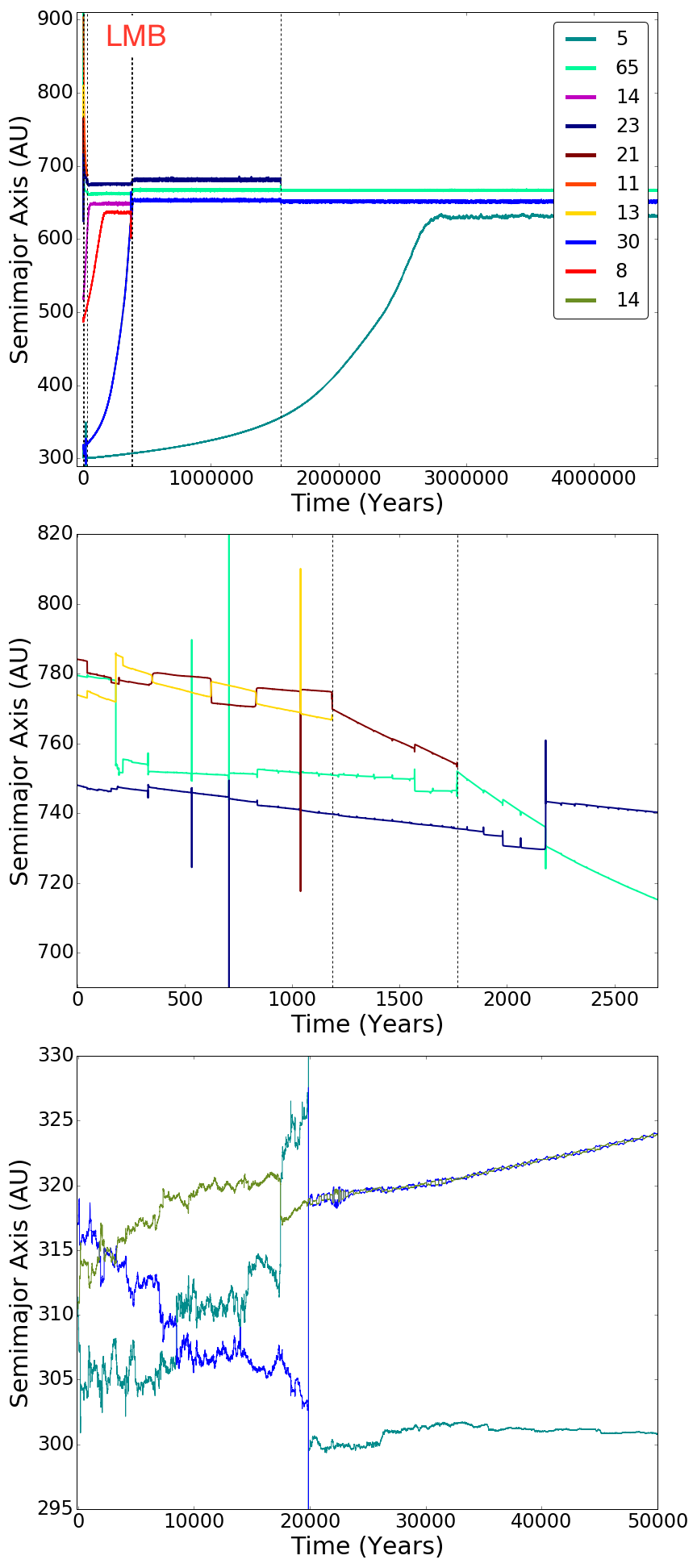}
\caption{Migration in the LMB run, with the same notation as Figure~\ref{fig:pathRLMA}. The top figure shows the first 4.5 Myr which is the period during which binary formation occurs, and all sBHs migrate towards the migration trap to stable orbits where they remain for the rest of the 10 Myr run. The middle figure is a zoomed in view of the first period of binary formation and the bottom figure is a zoomed in view of the interaction between the three innermost sBHs, two of which end up co-orbital.}
\label{fig:pathRLMB}
\end{figure}

\begin{figure}
\includegraphics[width=0.445\textwidth]{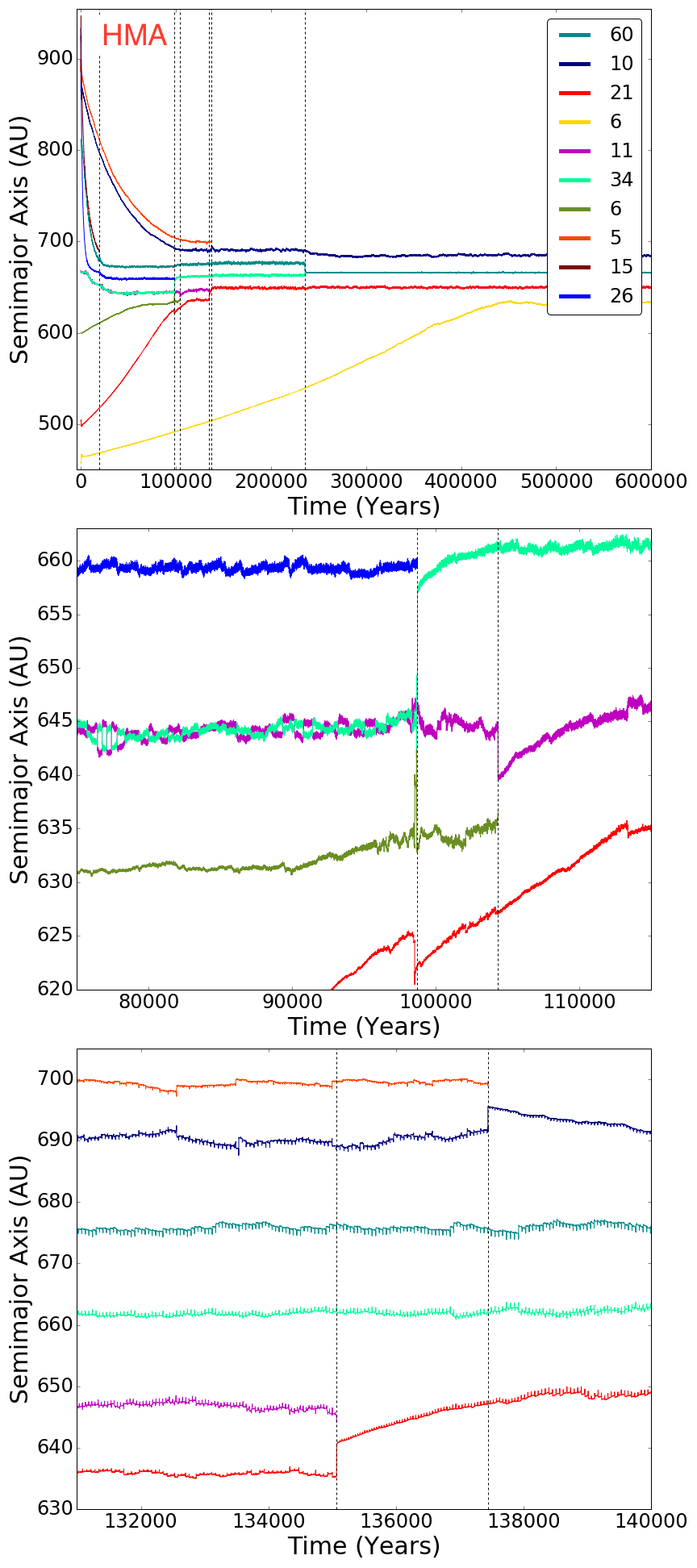}
\caption{Migration in the HMA run, with the same notation as Figure~\ref{fig:pathRLMA}. The top figure shows the first 600 kyr which is roughly the period during which binary formation occurs, and all sBHs migrate towards the migration trap to stable orbits where they remain for the rest of the 10 Myr run. 
The middle figure shows a zoomed in view of binary formation that breaks apart two co-orbital sBHs and the bottom panel shows a zoomed in view of a later period of binary formation. In the top panel the 6 $M_{\rm \sun}$ sBH is the last to reach the region of the migration trap, because it has a small initial semimajor axis. When it reaches the trap it ends up on its own resonant orbit, instead of merging with other sBHs.}
\label{fig:pathRHMA}
\end{figure}

\begin{figure}
\includegraphics[width=0.445\textwidth]{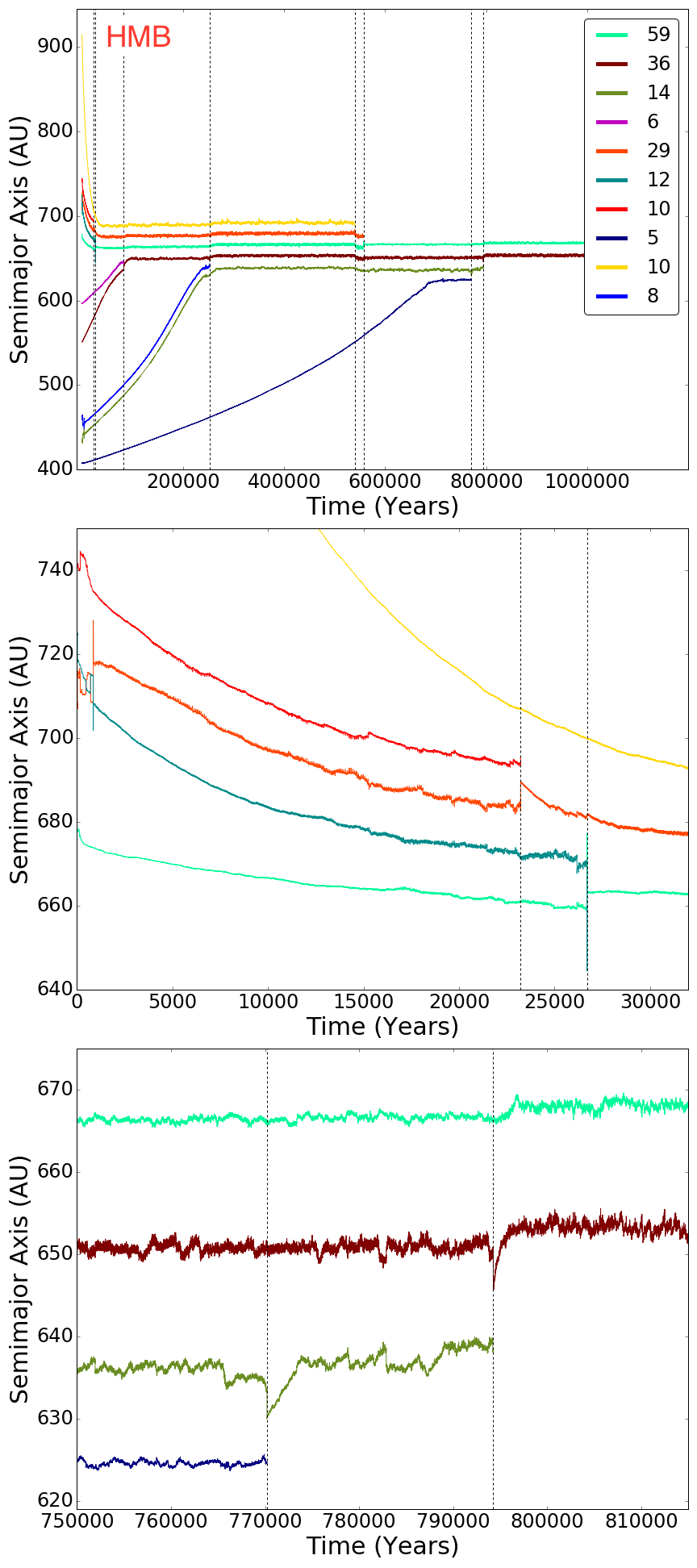}
\caption{Migration in the HMB run, with the same notation as Figure~\ref{fig:pathRLMA}. The top figure shows the first 1.1 Myr, which is roughly the period during which binary formation occurs, and all sBHs migrate towards the migration trap to stable orbits where they remain for the rest of the 10 Myr run. The bottom two figures are zoomed in views of the first (middle panel) and last (bottom panel) periods of binary formation. In the bottom panel the 5 $M_{\rm \sun}$ sBH that is the last to reach the migration trap region merges with a 9 $M_{\rm \sun}$ sBH that is on a resonant orbit with the other sBHs. This event breaks the resonance of the sBHs orbiting near the migration trap.}
\label{fig:pathRHMB}
\end{figure}

\begin{figure*}
\centering
\begin{tabular}{lccr}
\includegraphics[width=0.5\textwidth]{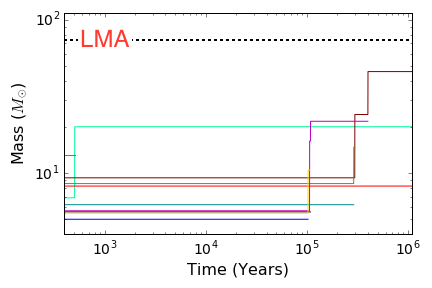} &
\includegraphics[width=0.5\textwidth]{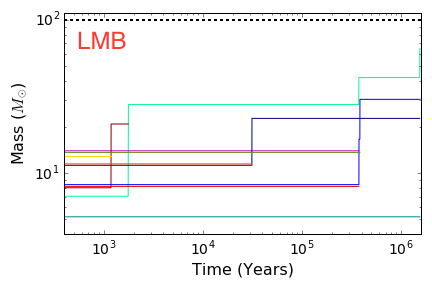} \\
\includegraphics[width=0.5\textwidth]{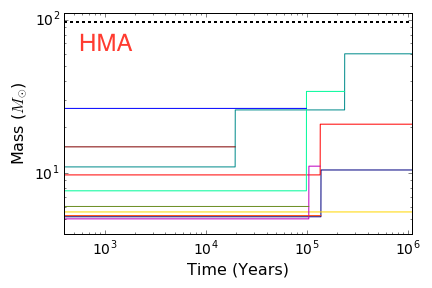} &
\includegraphics[width=0.5\textwidth]{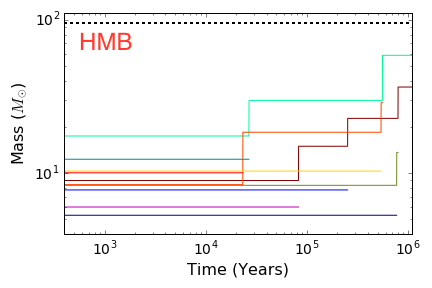}
\end{tabular}
\caption{The growth of sBHs through mergers over time for the LMA (top left), LMB (top right), HMA (bottom left), and HMB (bottom right) runs starting at 400 yr and ending at 2 Myr after which no mergers take place. Each colored line represents a sBH. The dashed black line represents the total mass of all sBHs in the model.}
\label{fig:growth}
\end{figure*}

\begin{figure}
\includegraphics[width=0.5\textwidth]{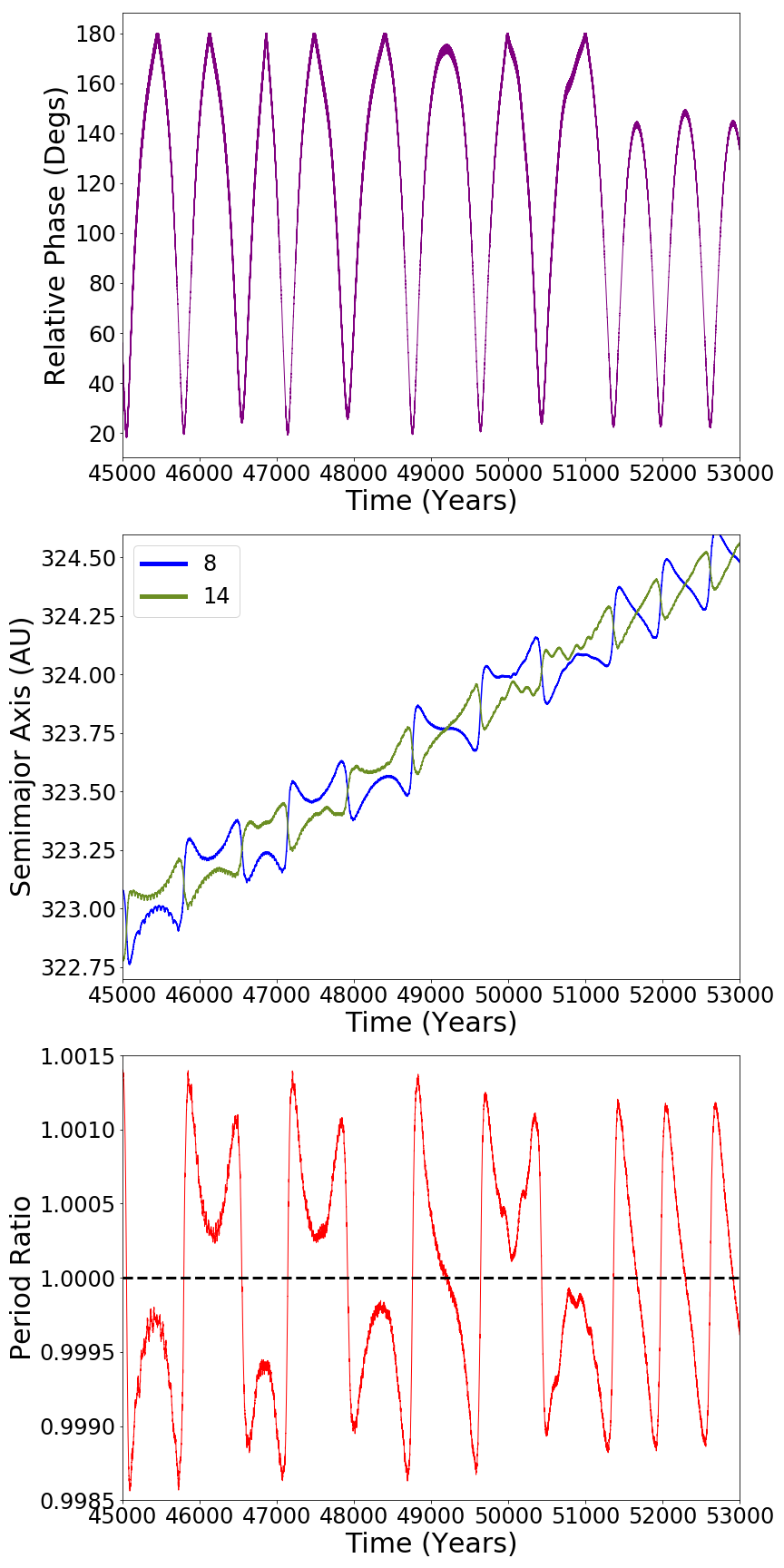}
\caption{Two sBHs from the LMB run in a stable horseshoe co-orbital configuration. 
The top panel shows the relative phase between the sBHs, the middle panel shows the semimajor axes of the two sBHs, which are labeled by their current masses, and the bottom panel shows the ratio of orbital periods around the SMBH.}
\label{fig:coorbit}
\end{figure}

\begin{figure*}[ht!]
\centering
\begin{tabular}{cc}
\includegraphics[width=0.3965\textwidth]{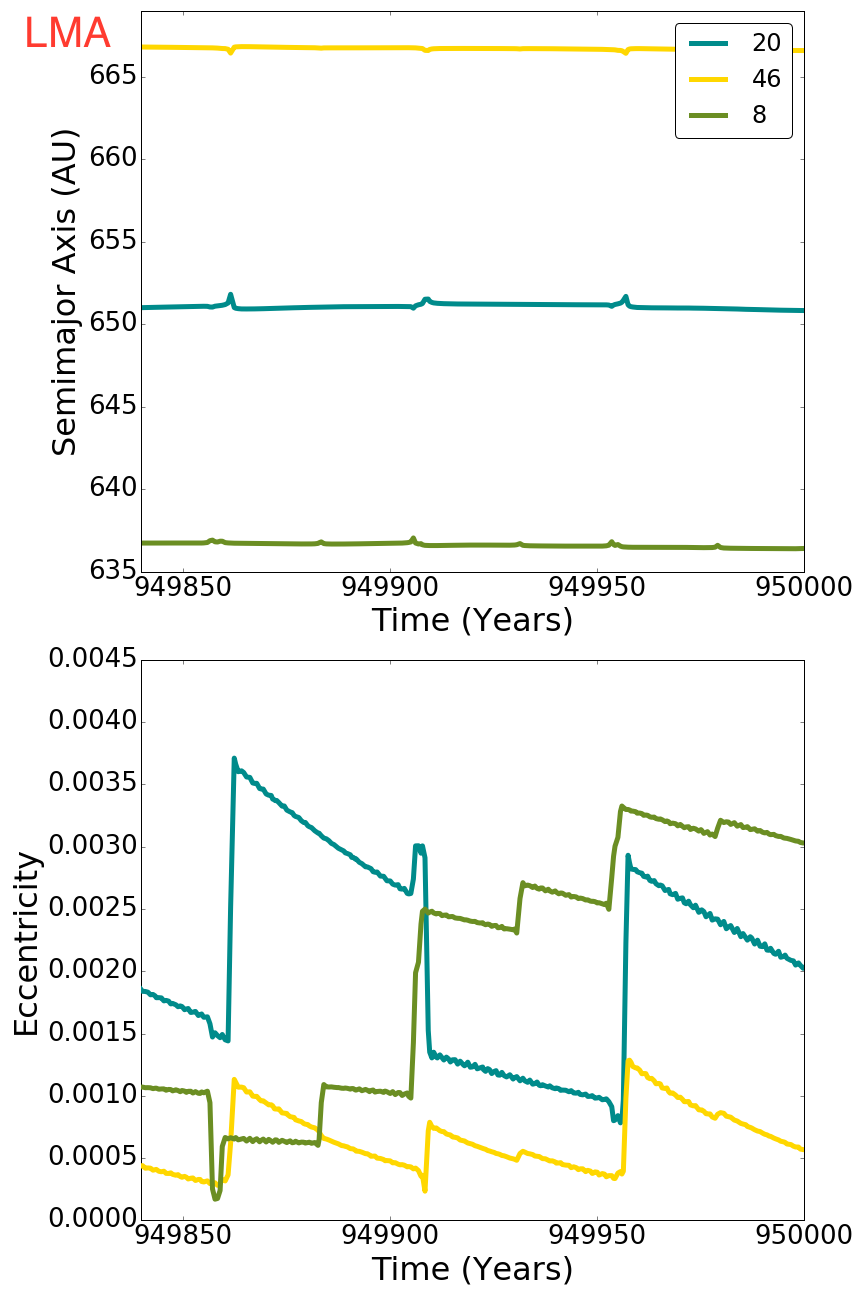} 
\includegraphics[width=0.3965\textwidth]{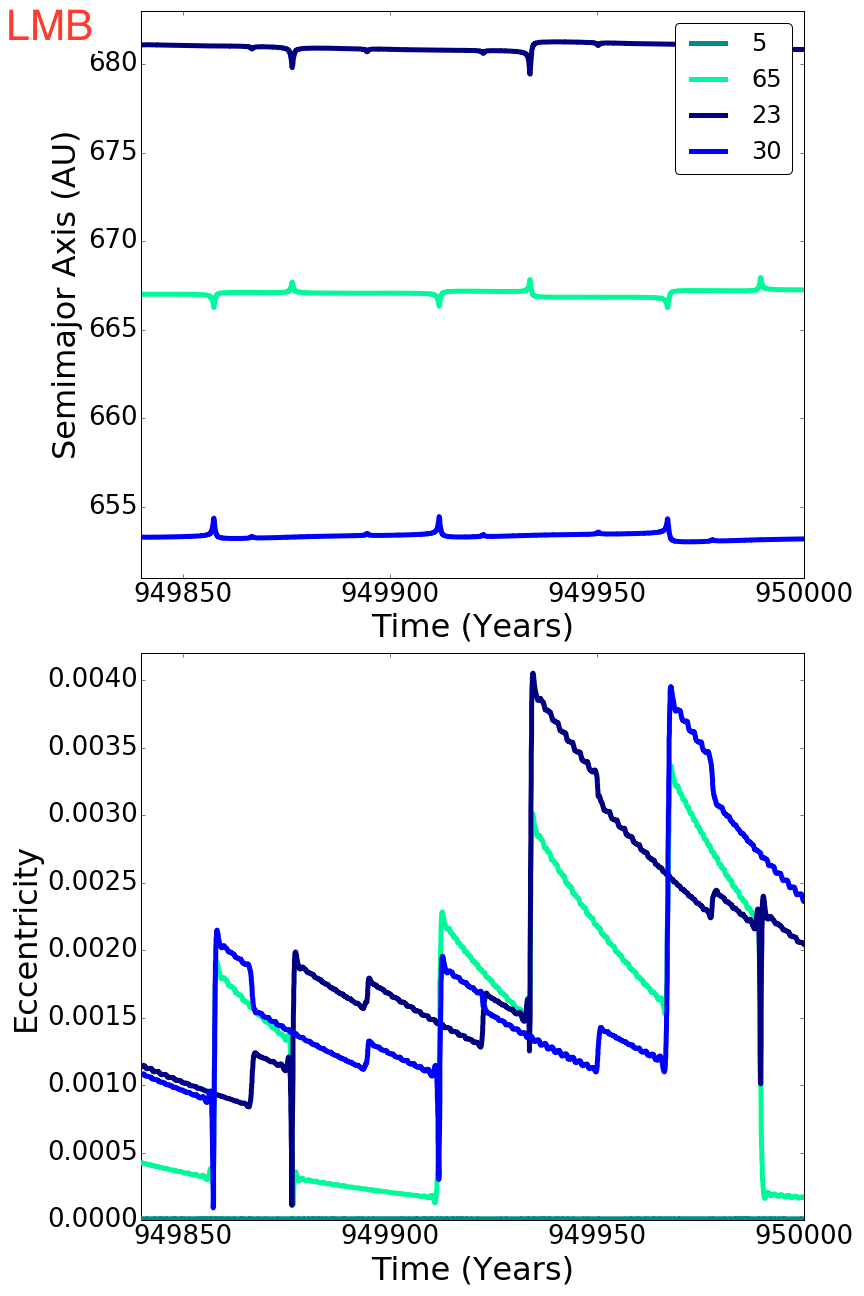}\\ 
\includegraphics[width=0.3965\textwidth]{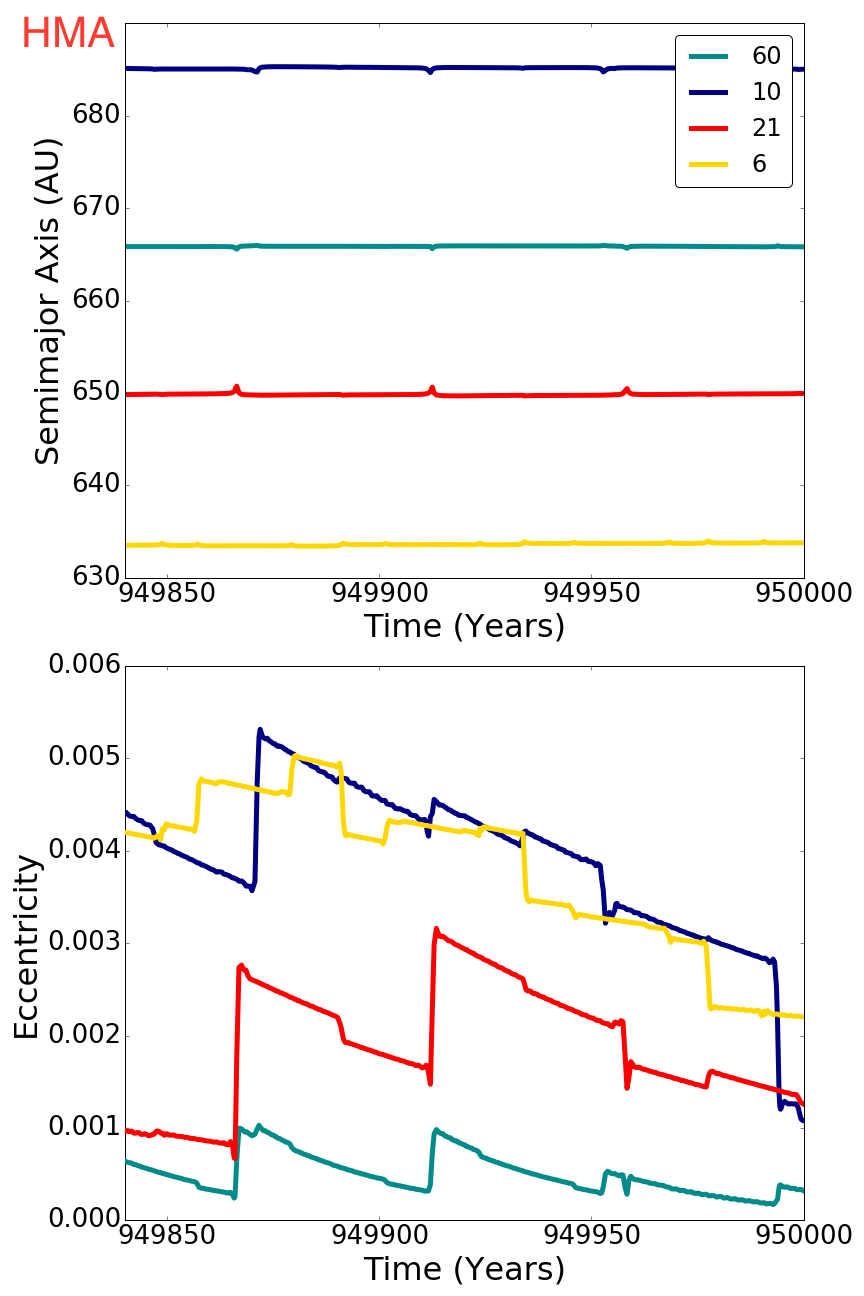}
\includegraphics[width=0.3965\textwidth]{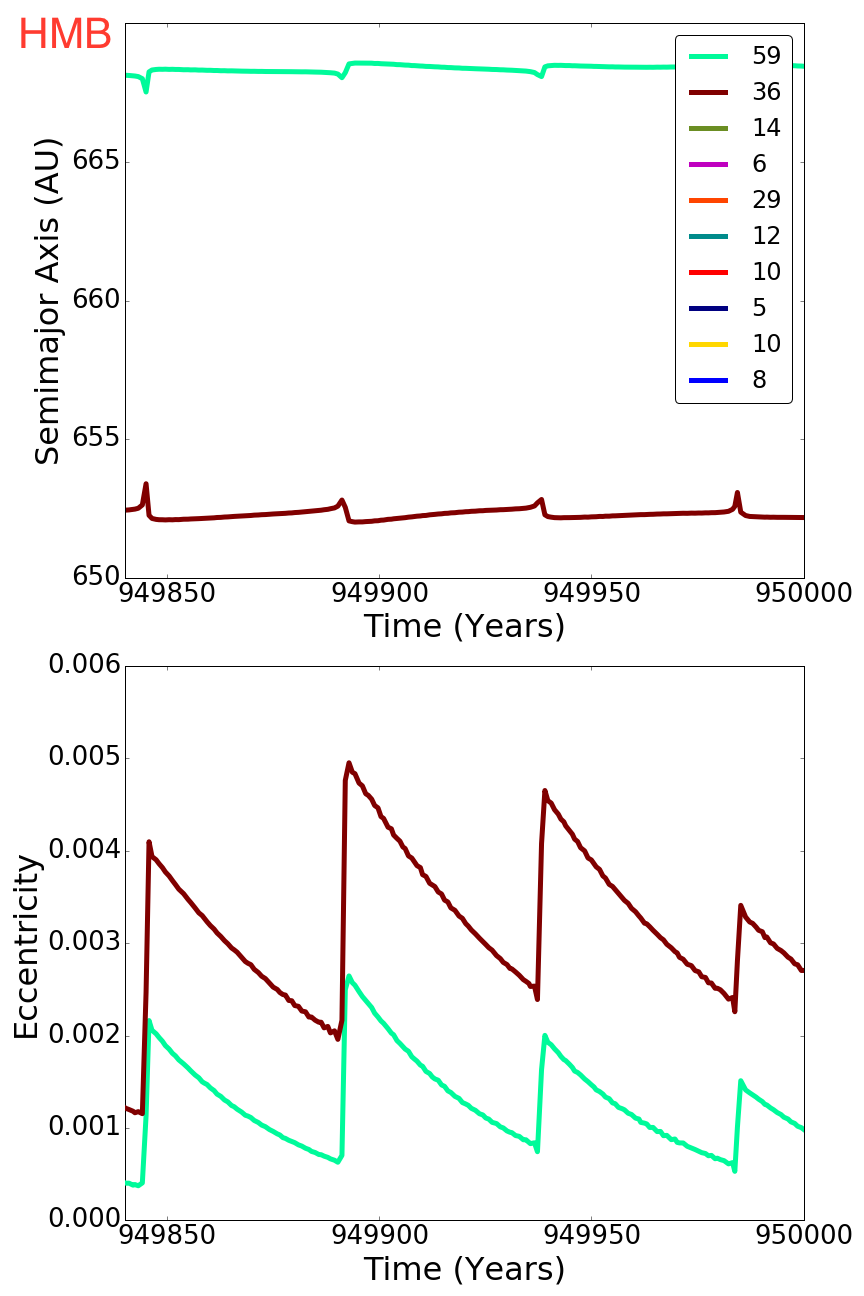}
\end{tabular}
\caption{Clockwise from the top left are zoomed in views of the stable resonant orbits of runs LMA, LMB, HMA, and HMB. The top figure for each run shows the semimajor axis and the bottom figure shows the eccentricity. In each plot each line represents one sBH and is labeled by its final mass in M$_{\rm \sun}$.}
\label{fig:MTZI}
\end{figure*}

\begin{figure}
\includegraphics[width=0.5\textwidth]{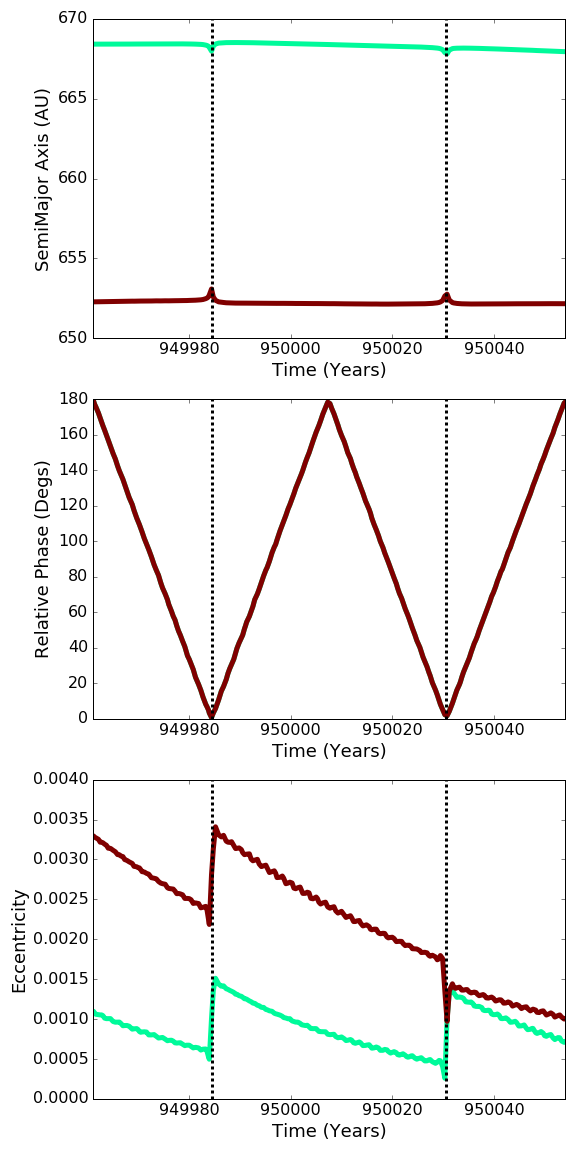}
\caption{A zoomed in view of the interactions between two sBHs from the HMB run on resonant orbits. The top plot shows the semimajor axis of the two sBHs, the middle plot shows the phase difference between the two sBHs, and the bottom plot shows the eccentricity of the two sBHs. When the phase difference reaches zero (represented by the vertical dashed line) the two sBHs are pulled towards each other by gravity making their orbits more eccentric. This eccentricity is then dampened until they are pulled towards each other again.}
\label{fig:phaseang}
\end{figure}

Figures \ref{fig:pathRLMA}--\ref{fig:pathRHMB} show the migration histories for runs LMA, LMB, HMA, and HMB, respectively (see Table~\ref{table:runs}). The top panel of all four figures shows the migration histories of the simulation up until all sBHs have reached stable orbits. In each simulation the sBHs remain at these final radii for the remainder of the 10 Myr run. The bottom two panels of these figures show examples of sBH interactions.

As the more massive sBHs migrate through the disk they overtake less massive sBHs and frequently form sBHBs. The time that elapses before the first binary capture of the simulation varies among the four runs from a few hundred years to roughly 20 kyr due to the randomly generated initial positions and eccentricities. Even if two sBHs have similar initial positions at the beginning of a simulation, if the orbits of the sBHs are too eccentric the relative kinetic energy of the two sBHs that approach each other within 1 $R_{\rm mH}$ may be higher than their binding energy preventing them from forming a sBHB (see Section \ref{sec:results:fiducial}).

Figure \ref{fig:growth} shows the build up in mass of the sBHs due to mergers for runs LMA (top left), LMB (top right), HMA (bottom left), and HMB (bottom right). In our simulation two sBHs are considered merged as soon as they form a sBHB (i.e. approach each other within 1 $R_{\rm mH}$; see Section \ref{sec:nbody}). 6-8 mergers occur in each run. The most massive sBH at the end of each run ranges from 45-65 $M_{\rm \sun}$, which represents 60-65$\%$ of the total mass of the run.

The time that elapses before all sBHs reach the migration trap also
varies and depends on the random generation of positions and
masses. The smaller the initial semimajor axis of a sBH the longer it
will take to migrate towards the trap, especially if it has a smaller
initial mass. 

Dynamical effects can produce some exceptions. For example,
in the LMB run (see Figure \ref{fig:pathRLMB}), there are three sBHs
with very small initial semimajor axes ranging from 310 AU to 320
AU. Being in such close initial proximity causes the sBHs to interact
with each other from the start, but they do not immediately 
form a sBHB. The least massive sBH only has a mass of 5 $M_{\rm \sun}$ and 
 after the three-body interaction
ends up on its own at approximately 300 AU. This low mass sBH left alone in a region with a very low migration rate takes nearly 3 Myr to finally make it to the migration trap. The two more massive sBHs (8 $M_{\rm \sun}$ and 14 $M_{\rm \sun}$) end up in a
stable horseshoe co-orbit as modeled by \cite{Cresswell_2006}, who found that it was common for planets in a protoplanetary disk to become co-orbital, occupying either horseshoe
or tadpole orbits that survived for the duration of their runs. Figure
\ref{fig:coorbit} shows the relative phase, semimajor axes, and ratio
of the orbital period around the SMBH for these two co-orbital
sBHs. Over a period of thousands of orbits the phase difference
between the two sBHs oscillates between 180$^o$ and 20$^o$. When the
phase difference is at a minimum the two sBHs swap radial
positions. Occasionally the migration rate of the more massive, 14
$M_{\rm \sun}$, sBH is large enough compared to the migration rate of the less massive, 8 $M_{\rm \sun}$, sBH that it overtakes it while the two are out of phase. However, the two sBHs still swap radial positions when they are closest to being in phase. As a result the 8 $M_{\rm \sun}$ sBH migrates at the rate of the 14 $M_{\rm \sun}$ sBH, which means the 8 $M_{\rm \sun}$ sBH reaches the migration trap at nearly double the rate it would alone.

In the HMA run, the cyan and purple sBHs in the center panel of Figure \ref{fig:pathRHMA} are also on a horseshoe co-orbit until the orbit is destabilized by the presence of a 26 $M_{\rm \sun}$ sBH, which the cyan sBH merges with. The co-orbital tadpole (i.e. Trojan) orbits that were observed by \cite{Cresswell_2006} are seen in runs F2 and F3 (see Figure \ref{fig:pathR_fid}).

In all cases, after several hundred kyr one sBH becomes massive enough
to dominate the region closest to the migration trap and lock all
other less massive sBHs in high-order
resonant orbits. sBHs migrating
towards the trap at later times will either merge with the sBHs
already populating 
resonant orbits (Figure \ref{fig:pathRHMB}), or end up on their own
resonant orbit (Figure \ref{fig:pathRHMA}). 

Figure \ref{fig:MTZI} shows the semimajor axes (upper panels) and eccentricities (lower panels) of the sBHs in or near the migration trap for the LMA (top left), LMB (top right), HMA (bottom left), and HMB (bottom right) runs. As in the F2 run (see Section \ref{sec:results:fiducial}), in the LMA, LMB, and HMB runs no sBH ends up exactly in the migration trap. Instead the most massive sBH ends up 1--2.5 AU from the migration trap, where it becomes locked in a resonant orbit  with the other sBHs. The semimajor axes of the sBHs around the SMBH spike periodically as the sBHs on resonant orbits exchange angular momentum with each other and the sBHs in the migration trap get pushed back into resonance. The sudden change in the orbit's semimajor axis causes a spike in eccentricity that is then dampened by the gas. Figure \ref{fig:phaseang} shows one example from the HMB run of these interactions of two sBHs on a 27:28 resonance. When the phase difference between the sBH in the migration trap and the sBH on a resonant orbit is zero, the two are pulled towards each other by their mutual gravitational attraction. This temporarily drives an increase in the eccentricity of their orbits, before it is gradually dampened once again by the gas disk.

These orbits remain stable for 9 Myr to the end of runs LMA, HMA, and HMB, suggesting that trapping sBHs in resonant orbits around a migration trap could prevent more massive sBHs from building up. However in the LMB run, as in the F3 run (see Section \ref{sec:results:fiducial}), a perturbative force caused by disk turbulence pushes the 23 $M_{\rm \sun}$ sBH out of resonance so that it merges with the 42 $M_{\rm \sun}$ sBH in the migration trap. Therefore disk turbulence could provide a mechanism to break resonances, and create more massive sBHs. \cite{horn} showed that increasing levels of disk turbulence makes this mechanism even more efficient. \cite{batygin} worked out an analytic solution for the breaking of resonances by turbulence for protoplanetary disks and found that the disruption of resonances by turbulence depends most strongly on the migrator-central mass ratio. For the migrator-central mass ratios and other relevant parameters in our simulations, their analytic solution agrees with our conclusion that turbulence could play a role in disrupting resonances.

The initial inclinations of the sBHs were very small, and all sBHs were quickly ground down into flat orbits in less than 50 yr. The initial eccentricities played a role in our models in preventing early sBHB formation, but were also a transient effect and were dampened by the gas in roughly 10 kyr. Larger initial values for inclination and eccentricity would likely delay sBHB formation because it would increase the relative kinetic energy of two sBHs. However, these larger inclinations and eccentricities will eventually be dampened by the gas disk, and as sBHs are ground down into the disk and their orbits are circularized, they would start to form sBHBs with other sBHs at later times.

\begin{figure}
\includegraphics[width=0.5\textwidth]{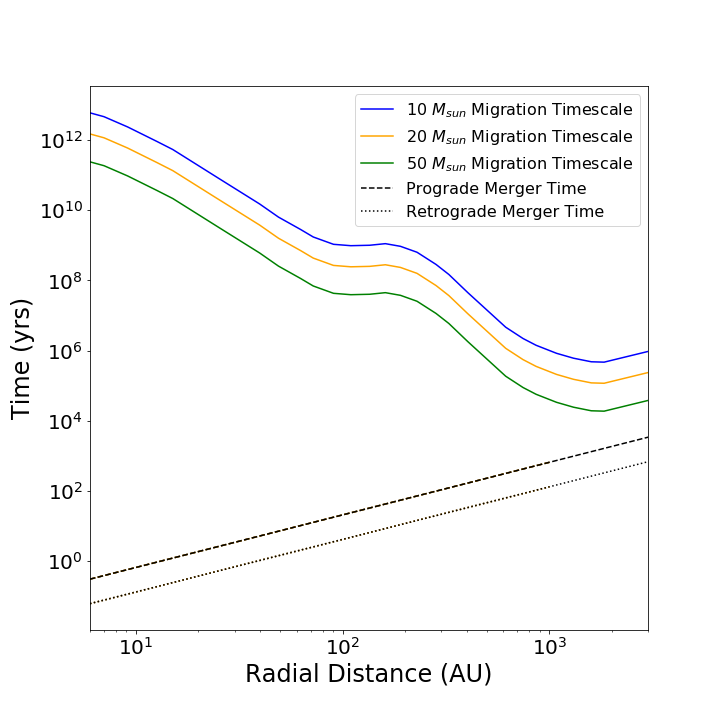}
\caption{Various timescales in years are plotted as a function of
  radial distance from the SMBH in AU. The blue, yellow, and green
  lines represent the approximate time for 10 $M_{\rm \sun}$, 20
  $M_{\rm \sun}$ and 50 $M_{\rm \sun}$ sBHs to migrate from their current location 
  to the SMBH due to only
  migration torques. The dashed and dotted black lines represent the merger
  time for a sBHB that forms when two sBHs are within a mutual Hill radius 
  (see Equation \ref{eq:rhill}) for a
  prograde orbiting sBHB and a retrograde orbiting sBHB,
  respectively. The merger timescales are significantly shorter than
  the migration timescales, suggesting that the probability of the
  sBHB failing to merge due to an encounter with a tertiary body is
  low. } 
 \label{fig:timescales}
\end{figure}

\section{Summary and Discussion}
\label{sec:discuss}

We have simulated the migration of compact objects in a model AGN disk
\citep{Sirko:2003aa}, using an analytic model developed from simulations of
the migration of protoplanets in protoplanetary disks. We have found that
migration due to gas torques in AGN disks can provide 
an efficient mechanism to create a population of hard compact object binaries
remarkably quickly, replicating the results of \cite{horn} for protoplanets in a protostellar disk, but for the case of sBHs in an AGN disk. 

\cite{mckernan18} parameterized the rate of sBH-sBH mergers in AGN disks as,
\begin{equation}
\label{eq:rate}
\begin{multlined}
R = 12 \mbox{ Gpc}^{-3}\mbox{
  yr}^{-1}\frac{N_{\rm GN}}{0.006\mbox{ Mpc}^{-3}}\frac{N_{\rm BH}}{\num{2e4}}\frac{f_{\rm AGN}}{0.1} \\
   \mbox{X } \frac{f_{\rm d}}{0.1}\frac{f_{\rm b}}{0.1}\frac{\epsilon}{1}\left(\frac{\tau_{\rm AGN}}{10
  \mbox{ Myr}}\right)^{-1},
  \end{multlined}
\end{equation}
where $N_{\rm BH}$ is the number of sBHs in an AGN disk, $N_{\rm GN}$
is the average number density of galactic nuclei in the Universe,
$f_{\rm AGN}$ is the fraction of galactic nuclei with AGN that last for
time $\tau_{\rm AGN}$, $f_{\rm d}$ is the fraction of sBHs that end up in the AGN disk, $f_{\rm b}$ is the fraction of sBHs that form binaries, and
$\epsilon$ represents the fractional change in $N_{\rm BH}$ over one full
AGN duty cycle. Using our finding that within the inner 1000~AU of an
AGN disk 60--80\% of sBHs form sBHBs in the lifetime of our AGN disk,
we can use 0.6--0.8 as an upper limit on $f_{\rm b}$, giving an upper
limit on the merger rate of 72 Gpc$^{-3}$~yr$^{-1}$. 
This value is an upper limit because, although our model assumes a uniform distribution of sBHs throughout the disk, sBHs in the outer disk, further from the migration trap, may merge less frequently. In addition, this upper limit assumes that sBHs orbiting in the retrograde direction would have similar sBHB formation rates, which is unlikely because migration torques on retrograde orbiters should be much weaker. We defer a more realistic prediction of retrograde orbiter merger rates and merger rates of sBHs in the outer disk to future work.

Uncertainties in AGN disk structure result in a wide variety of plausible theoretical models to describe these disks. However, migration traps should occur in any disk where there is a rapid change in the surface density gradient \citep{bellovary}. Such rapid changes are likely to occur in most actual disks, since radiation pressure is expected to inflate the inner disk. This paper is intended to highlight the qualitative behavior of objects at the migration trap. Regardless of the location of these migration traps, whether they are at 331~$R_{\rm s}$ as in the \citet{Sirko:2003aa} model or about 225~$R_{\rm s}$ as in the \citet{thompson} model \citep{bellovary}, most of the binary formation will take place in the immediate vicinity of the migration traps. We expect the qualitative behavior around the migration trap to be similar regardless of AGN disk model, although having a migration trap at a different distance from the SMBH as in \citet{thompson} will affect how long it will take sBHs to migrate to the trap. Additionally, different disk models have different surface densities, which will affect migration rates. If these surface densities are lower than in \citet{Sirko:2003aa}, as they are in  \citet{thompson}, the migration rates will be lower. We defer simulations of migration in alternative AGN disk models to future work.

We highlight that although we have taken the compact objects to be
sBHs here, similar results apply to any objects embedded in the
AGN disk, including neutron stars, white dwarfs, or
main sequence or evolved stars, although their typically lower masses will result in slower migration rates. Our demonstration of how quickly binaries can form in AGN disks may help us to understand the behavior of other objects embedded in AGN disks. For example, \cite{davies} attributed the observed lack of red giant stars in the galactic center to direct collisions during single-binary encounters. We might suggest a simple alternative, albeit analogous mechanism motivated by our results in this paper:  main-sequence turn-off stars efficiently form (or are exchanged into) compact binaries, such that they form common envelope binaries (or some other variation of the myriad of possible binary evolution pathways) when the turn-off star evolves up the giant branch, preventing it from evolving in to a normal red giant star.  In short, a myriad of binary and stellar exotica could form in AGN disks.  These additional compact objects could also contribute non-negligibly to subsequent binary mergers and interactions \citep{leigh2016}, and even produce exotic populations that might contribute to the total light distribution in galactic nuclei non-negligibly, once the gas disk has dissipated and the SMBH is no longer actively accreting at high rates.

One assumption of our model is that sBHBs merge as soon as they form.  These binaries actually harden due to gas torques on a timescale that depends on the distribution of gas in the Hill sphere of the binary, and which also involves the complicated effects of accretion onto the sBHB and the resulting feedback. We justify our assumption by comparing the migration timescale to the binary hardening time scale. \citet{baruteau2011} modeled the hardening
of binaries in a gas disk. Their models showed that it takes roughly
1000 orbits of binary stars around the binary's center of mass
to reduce the semimajor axis of the binary by a factor of two if the
binary is rotating in the prograde direction with respect
to its orbit around the central mass, and only 200 orbits for retrograde rotation.

We assume that after the binary's semimajor axis has been halved 20
times, the sBHB separation is small enough that gravitational
radiation will rapidly merge the sBHB to form a single sBH of mass
$m_{\rm i} + m_{\rm j}$. The binary inspiral time due to gravitational
wave emission alone \citep{peters}, neglecting any gas hardening
effects, exceeds the binary hardening timescale of 
4--200 $\times 10^3$ orbits as long as the binary eccentricity $e<0.9995$. Note that this estimate may be a significant underestimate of the actual time to merger, since gas hardening may become less efficient as the binary shrinks. However, we have also neglected the possibility of hardening encounters due to tertiary objects in the disk, which will accelerate the rate of binary hardening \citep{leigh2016,leigh}. Both of these complications will require further study in future work.

Given our assumptions, Figure \ref{fig:timescales} shows the approximate radial dependence of the timescales of mergers for sBHBs rotating in prograde and retrograde directions, for sBHBs orbiting in the prograde direction through the disk. In our simulations these timescales will be equivalent for all mass sBHBs because sBHs are considered to form a sBHB when they approach each other within a mutual Hill radius, which is $\propto(m_{\rm i} + m_{\rm j})^{1/3}$.
For comparison, in Figure~\ref{fig:timescales} we plot the time for 10 $M_{\rm \sun}$, 20 $M_{\rm \sun}$ and 50 $M_{\rm \sun}$ sBHs to migrate from their current radial location to the SMBH due to migration torques. Recall that the migration torques vary as a function of radius and temperature, surface density and disk aspect ratio at each radius. Since the migration timescales of these objects are at least an order of magnitude larger than the time it would take for a sBHB of the same mass to merge, we can see that the likelihood of a tertiary encounter from another sBH is low.

This low likelihood is important because while a tertiary encounter could accelerate a binary merger \citep{leigh}, the third sBH would be ejected in the process, and because the sBHB will already be merged in our simulation, it is not possible for a third body to gain energy from a three-body encounter. However, our simulations do permit binary formation to occur via three-body interactions in a limited set of realistic circumstances. That is, three initially isolated sBHs could end up in a sufficiently small volume that their mutual gravitational attraction dominates locally, and a chaotic three-body interaction ensues.  If one star is ejected, the other two remaining sBHs could form a binary.  

The dissipative effects of the gas actually
enhance the probability of such three-body mediated 
binary formation occurring. 
The critical orbital separation of a sBHB for which the kinetic energy
of a third isolated sBH is equal to the orbital energy of the sBHB is
known as the hard-soft boundary. Third body encounters with hard binaries promote hardening, while with soft binaries they can promote ionization. 
In an AGN disk the hard-soft boundary 
for a sBHB in a circular orbit is \citep{leigh}
\begin{equation}
\label{eq:hs_radius}
a_{\rm HS,disk}=(12)^{1/3}R_{\rm H}(\mu_{\rm b}/M_{\rm 3})^{1/3},
\end{equation}
where $R_{\rm H}$ is the Hill radius, $\mu_{\rm b}$ is the reduced mass
$M_{\rm 1}M_{\rm 2}/(M_{\rm 1}+M_{\rm 2})$ of the binary, and $M_{\rm 3}$ is the
mass of the 
third sBH. Since we consider sBHBs to be merged once they are within a Hill radius, as long as $12 \mu_{\rm b} / M_{\rm 3} > 1$, the kinetic energy from a prograde tertiary sBH should not be enough to ionize a sBHB in our simulations.    

Looking at examples in our simulations of sBHBs that have a close
encounter with a third sBH on timescales shorter than the merger
timescales in Figure \ref{fig:timescales}, we find only one instance
where a third sBH is massive enough that $a_{\rm HS,disk} < R_{\rm H}$.
However in this case the third sBH never approaches closer than 
10 $R_{\rm H}$ from the sBHB, making it too distant to ionize the
sBHB. Therefore in our models the ionization of our binaries by a
prograde third body interaction appears to be rare. 

Finally it is possible for one of two sBHs within a mutual Hill radius of each other to be ejected, even if the binding energy of the two sBHs is less than their relative kinetic energy and there is no tertiary interaction. Preliminary results from Secunda et al.\ (in prep.) suggest that this is rare. In Secunda et al.\ (in prep.) the merger boundary is reset to 0.65~$R_{\rm{mH}}$ for the same runs as in this paper. This boundary was chosen to allow us to study some of the properties of the sBHBs we were forming without requiring unreasonably large computational resources, in the form of integration time. In 4 out of 32 cases, sBHs that would have merged under the criteria presented in this paper did not merge when the boundary for merger is 0.65~$R_{\rm{mH}}$. Instead these sBHs swapped orbits. This orbit swapping in place of sBHB formation is already seen in runs with the sBHB formation criterion set to 1 $R_{\rm{mH}}$ (see the yellow and brown lines in the center panel of Figure \ref{fig:pathRLMB}). Additionally, those sBHs that failed to merge initially later were able to merge with other sBHs. Therefore, the merger histories of runs with a more stringent merger criterion were qualitatively identical to those presented above. 

Future work that includes the relevant gas physics should evolve sBHBs to much smaller merger boundaries of 0.1 to 0.01~$R_{\rm{mH}}$ to further probe the poorly understood evolution of sBHBs in a gas disk. We do not do so here for simplicity's sake, since whether a sBHB in a gaseous accretion disk will be able to merge is still an open question. For example, recent work by \cite{moody} found that the circumbinary disk around BHBs can actually exert a net positive torque on the BHB, causing its semi-major axis to increase. The properties of the sBHBs formed in our simulations should serve as useful, physically motivated inputs for future hydrodynamic simulations of BHB evolution in gas disks.

Our model is efficient at building up massive sBHBs on timescales far
shorter than the lifetime of the AGN disks that host them. We argue
that these sBHBs are likely to merge, producing gravitational wave
events such as those observed by LIGO. However, future work using
hydrodynamic simulations is needed to better describe the interactions
between the gas disk and the sBHs, in particular examining the binary
hardening timescale due to gas torques. More work is also needed to
model the evolution of the sBH population as additional compact
objects either drift inward or have their orbital inclination ground
down into the inner region of the disk where they may be able to break
resonances and form additional sBHBs and more massive sBHs. We have
also completely ignored the role of retrograde orbiters in this paper, and the population of objects on retrograde orbits that 
can ionize binaries embedded in the disk. General relativistic effects are also not included in our model. Ultimately a full, three-dimensional, time-evolving AGN disk model should be used to provide the most accurate predictions for the merger rates of sBHs and the build-up of over-massive sBHs.  In the meantime, constraints from the next few LIGO runs on mergers from this model channel should help put limits on models of AGN disks (in particular, the presence or absence of density gradients likely to produce migration traps), such as those used here.

\acknowledgments
We thank BridgeUP: STEM Brown Scholars Juliette Cornelis, Denelis
Ferreira, Ariba Khan, Anna Li, Audrey Soo, and Anay Vicente for their
test runs and preliminary figures, and the anonymous referee for a useful report that improved the
   clarity and accuracy of the paper. 
 A.S. was supported by a fellowship
from the Helen Gurley Brown Revocable Trust, and N.W.C.L was supported by a Kalbfleisch Fellowship, both at the
American Museum of Natural History. 
M.-M.M.L. was partly 
funded by NASA Astrophysical Theory Grant NNX14AP27G and NSF Grant AST18-15461. 
N.W.C.L. and M.-M.M.L. were partly funded by NSF Grant
AST11-0395. B.M. and K.E.S.F. was partly supported by NSF PAARE
AST11-53335 and NSF PHY11-25915.  J.M.B. acknowledges support from
PSC-CUNY award 60303-00 48, and W.L. acknowledges support from Space
Telescope Science Institute through grant HST-AR-14572 and the NASA
Exoplanet Research Program through grant 16-XRP16~2-0065

\appendix

\section{Dynamical Torque}
\label{A}

As we note in Section \ref{sec:intro}, the underlying physics of migration remains uncertain. For example, \cite{paardekooper2014} found that for low enough viscosity, if there is a radial gradient in vortensity
the planets in a protoplanetary disk can experience dynamical torques in
addition to the static torques implemented in our model. These
dynamical torques could act to slow down inward migration and even
lead to runaway outward migration. However, at least in the particular example we focus on in this paper, these effects do not necessarily act.  \citet{paardekooper2014} emphasizes that dynamic migration only sets in when 
    \begin{equation} 
    \label{eq:dynamic} 
    k \sim m_c \tau_\nu / \tau_{\rm mig} >1/2, 
    \end{equation}
where the coorbital gas mass in planet masses is 
    \begin{equation} 
    m_c = 4 q_d \tilde{x}_s  / q, 
    \end{equation} 
the migration time scale 
    \begin{equation}
    \tau_{\rm mig} = \frac{\pi}{2} \frac{h^2}{q_d q\Omega},
    \end{equation}
and the time for viscosity to adapt the co-orbital vortensity to the
ambient value is
    \begin{equation}
    \tau_{\nu} = x_s^2 / \nu.
    \end{equation}
The half-width of
the horseshoe region, in units of the planet's orbital radius $r_p$,
is $\tilde{x}_s \simeq (q/h)^{1/2}$ \citep{paardekooper2009}.

In our example, with central SMBH mass $M = 10^8 M_{\odot}$, the mass ratio of the disk to the central SMBH $q_d = 0.37$, the mass ratio of the
orbiter to the SMBH, $q = 1$--3$\times 10^{-7}$ (Tab.~\ref{table:runs}), the angular velocity of the orbiter
$\Omega = (GM/r^3)^{1/2}$, the Shakura-Sunyaev
(\citeyear{shakura_sunyaev}) 
viscosity parameter $\alpha = c_s H / \nu = 0.01$ \citep{Sirko:2003aa},
and, at the trap radius $r = 3.2 \times 10^{-3}$~pc (Sect.~3), the disk aspect ratio
$h = H/r = 0.05$ (see Fig.~1), and the sound speed $c_s \simeq 10^{7}
\mbox{ cm s}^{-1}$ \citep[Fig.~2,][including both radiation and
thermal pressure]{Sirko:2003aa}.  We can derive 
    \begin{equation}
    k = \frac{8}{\pi} \left(\frac{G M}{r}\right)^{1/2} \frac{q^{3/2}q_d^2}{\alpha c_s h^{9/2}}. 
    \end{equation}
In our case, we find $k = 0.09$--0.27, satisfying the condition (Eq.~\ref{eq:dynamic}) that dynamical migration be ineffective.  As our objects grow by merger, this will eventually no longer be true, though, so future work will need to include this effect.

\bibliographystyle{aasjournal}

\begin{thebibliography}{}
\expandafter\ifx\csname natexlab\endcsname\relax\def\natexlab#1{#1}\fi
\providecommand{\url}[1]{\href{#1}{#1}}

\bibitem[{{Abbott} {et~al.}(2016){Abbott}, {Abbott}, {Abbott}, {Abernathy},
  {Acernese}, {Ackley}, {Adams}, {Adams}, {Addesso}, {Adhikari}, \&
  et~al.}]{abbott}
{Abbott}, B.~P., {Abbott}, R., {Abbott}, T.~D., {et~al.} 2016, \apjl, 833, L1

\bibitem[{{Alexander} \& {Ferguson}(1994)}]{alexander_ferguson}
{Alexander}, D.~R., \& {Ferguson}, J.~W. 1994, \apj, 437, 879

\bibitem[{{Antonini}(2014)}]{antonini}
{Antonini}, F. 2014, \apj, 794, 106

\bibitem[{{Antonini} \& {Rasio}(2016)}]{Antonini_rasio}
{Antonini}, F., \& {Rasio}, F.~A. 2016, \apj, 831, 187

\bibitem[{{Bahcall} \& {Wolf}(1976)}]{bahcall_wolf}
{Bahcall}, J.~N., \& {Wolf}, R.~A. 1976, \apj, 209, 214

\bibitem[{{Bai} \& {Stone}(2010)}]{bai_stone} {Bai}, X.-N., \& {Stone}, 
J.~M. 2010, \apjl, 722, L220

\bibitem[{{Baruteau} {et~al.}(2011){Baruteau}, {Cuadra}, \&
  {Lin}}]{baruteau2011}
{Baruteau}, C., {Cuadra}, J., \& {Lin}, D.~N.~C. 2011, \apj, 726, 28

\bibitem[{{Baruteau} \& {Lin}(2010)}]{baruteau_lin}
{Baruteau}, C., \& {Lin}, D.~N.~C. 2010, \apj, 709, 759

\bibitem[{{Batygin} \& {Adams}(2017)}]{batygin} {Batygin}, K., \& {Adams}, 
F. C. 2017, \apj, 153, 120

\bibitem[{{Belczynski} {et~al.}(2016){Belczynski}, {Holz}, {Bulik}, \&
  {O'Shaughnessy}}]{Belczynski}
{Belczynski}, K. and {Holz}, D.~E. and {Bulik}, T. \& {O'Shaughnessy}, R. 2016,
  \apjl, 819, L17

\bibitem[{{Bellovary} {et~al.}(2016){Bellovary}, {Mac Low}, {McKernan}, \&
  {Ford}}]{bellovary}
{Bellovary}, J.~M., {Mac Low}, M.-M., {McKernan}, B., \& {Ford}, K.~E.~S. 2016,
  \apjl, 819, L17

\bibitem[{{Ben{\'i}tez-Llambay} {et~al.}(2015){Ben{\'i}tez-Llambay}, {Masset}, \& {Koenigsberger}}]{b-llambay}
{Ben{\'i}tez-Llambay}, P., {Masset}, F., \& {Koenigsberger}, G. 2015,
  Nature, 520, 63
  
\bibitem[{{Blaes} \& {Balbus}(1994)}]{blaes} {Blaes}, O.~M., \& {Balbus}, 
S.~A. 1994, \apj, 421, 163

\bibitem[{Coleman \& Nelson(2014)}]{coleman}
Coleman, G.~A.~L., \& Nelson, R.~P. 2014, \mnras, 445, 479

\bibitem[{{Comins} {et~al.}(2016){Comins}, {Ramanova}, {Koldoba}, {Ustyugova}, {Blinova}, \& {Lovelace}}]{comins}
{Comins}, M.~L., {Ramanova}, M.~M., {Koldoba}, A.~V., {Ustyugova}, G.~V., {Blinova}, A.~A., \& {Lovelace}, R.~V.~E. 2016,
  \mnras, 459, 3482

\bibitem[{Cresswell \& Nelson(2006)}]{Cresswell_2006}
Cresswell, P., \& Nelson, R.~P. 2006, \aap, 450, 833
\newblock \url{http://dx.doi.org/10.1051/0004-6361:20054551}

\bibitem[{{Cresswell} \& {Nelson}(2008)}]{cresswell_nelson}
{Cresswell}, P., \& {Nelson}, R.~P. 2008, \aap, 482, 677

\bibitem[{{Davies} {et~al.}(1998){Davies}, {Blackwell}, {Bailey}, \&
  {Sigurdsson}}]{davies}
{Davies}, M.~B., {Blackwell}, R., {Bailey}, V.~C., \& {Sigurdsson}, S. 1998,
  \mnras, 301, 745

\bibitem[{{Davis} {et~al.}(2010){Davis}, {Stone}, \& {Pessah}}]{davis}
{Davis}, S.~W., {Stone}, J.~M., \& {Pessah}, M.~E. 2010, \apj, 713, 52

\bibitem[{{Dittkrist} {et~al.}(2014){Dittkrist}, {Mordasini}, {Klahr}, {Alibert}, \& {Henning}}]{dittkrist}
{Dittkrist}, K.~M., {Mordasini}, C., {Klahr}, H., {Alibert}, Y. \& {Henning}, T. 2014, \aap, 567, A121

\bibitem[{{Eklund} \& {Masset}(2017)}]{eklund_masset}
{Eklund}, H., \& {Masset}, F.~S. 2017, \mnras, 469, 206

\bibitem[{{Fryer} {et~al.}(2012){Fryer}, {Belczynski}, {Wiktorowicz},
  {Dominik}, {Kalogera}, \& {Holz}}]{fryer}
{Fryer}, C.~L., {Belczynski}, K., {Wiktorowicz}, G., {et~al.} 2012, Nature, 534,
  512

\bibitem[{{Gammie}(1996)}]{gammie}
{Gammie}, C.~F. 1996, \apj, 457, 355

\bibitem[{{Goldreich} \& {Tremaine}(1979)}]{goldreich_tremaine}
{Goldreich}, P., \& {Tremaine}, S. 1979, \apj, 233, 857

\bibitem[{{Guilet} {et~al.}(2013){Guilet}, {Baruteau}, \& {Papaloizou}}]{guilet}
{Guilet}, J., {Baruteau}, C. \& {Papaloizou}, J.~C.~B. 2013, \mnras, 430,
  1764

\bibitem[{{Haehnelt} \& {Rees}(1993)}]{haehnelt}
{Haehnelt}, M.~G., \& {Rees}, M.~J. 1993, \mnras, 263, 168

\bibitem[{Hailey {et~al.}(2018)Hailey, Mori, Bauer, Berkowitz, Hong, \&
  Hord}]{Hailey_2018}
Hailey, C.~J., Mori, K., Bauer, F.~E., {et~al.} 2018, Nature, 556, 70.
\newblock \url{http://dx.doi.org/10.1038/nature25029}

\bibitem[{{Hellary} \& {Nelson}(2012)}]{hellary_nelson}
{Hellary}, P., \& {Nelson}, R. P. 2012, \mnras, 419, 2737

\bibitem[{Hopman \& Alexander(2006)}]{Hopman:2006aa}
Hopman, C., \& Alexander, T. 2006, Astrophys.J., 645, L133.
\newblock \url{https://arxiv.org/abs/astro-ph/0603324}

\bibitem[{{Horn} {et~al.}(2012){Horn}, {Lyra}, {Mac Low}, \&
  {S{\'a}ndor}}]{horn}
{Horn}, B., {Lyra}, W., {Mac Low}, M.-M., \& {S{\'a}ndor}, Z. 2012, \apj, 750,
  34

\bibitem[{{Hubeny}(1990)}]{hubeny}
{Hubeny}, I. 1990, \apj, 351, 632

\bibitem[{{Iglesias} \& {Rogers}(1996)}]{iglesias_rogers}
{Iglesias}, C.~A., \& {Rogers}, F.~J. 1996, \apj, 464, 943

\bibitem[{{Jiang} {et~al.}(2017){Jiang}, {Stone}, \& {Davis}}]{jiang_stone}
{Jiang}, Y.-F., {Stone}, J., \& {Davis}, S.~W. 2017, ArXiv e-prints,
  arXiv:1709.02845
  
\bibitem[{{Jim\'{e}nez} \& {Masset}(2017)}]{jimenez_masset}
{Jim\'{e}nez}, M.~A., \& {Masset}, F.~S. 2017, \mnras, 471, 4917

\bibitem[{{King} \& {Nixon}(2015)}]{king_nixon}
{King}, A., \& {Nixon}, C. 2015, \mnras, 453, L46

\bibitem[{{Kley} \& {Crida}(2008)}]{kley_crida}
{Kley}, W., \& {Crida}, A. 2008, \aap, 487, L9

\bibitem[{{Kroupa}(2002)}]{kroupa}
{Kroupa}, P. 2002, Science, 295, 82

\bibitem[{Laughlin \& Bodenheimer(1994)}]{Laughlin_1994}
Laughlin, G., \& Bodenheimer, P. 1994, The Astrophysical Journal, 436, 335.
\newblock \url{http://dx.doi.org/10.1086/174909}

\bibitem[{{Leigh} {et~al.}(2016){Leigh}, {Antonini}, {Stone}, {Shara}, \&
  {Merritt}}]{leigh2016}
{Leigh}, N.~W.~C., {Antonini}, F., {Stone}, N.~C., {Shara}, M.~M., \&
  {Merritt}, D. 2016, \mnras, 463, 1605

\bibitem[{{Leigh} {et~al.}(2018){Leigh}, {Geller}, {McKernan}, {Ford}, {Mac
  Low}, {Bellovary}, {Haiman}, {Lyra}, {Samsing}, {O'Dowd}, {Kocsis}, \&
  {Endlich}}]{leigh}
{Leigh}, N.~W.~C., {Geller}, A.~M., {McKernan}, B., {et~al.} 2018, \mnras, 474,
  5672
  
  \bibitem[{{Lesur} {et~al.}(2014){Lesur}, {Kunz}, \&
  {Fromang}}]{lesur}
{Lesur}, G., {Kunz}, M.~W., \&
  {Fromang}, S. 2014, \aap, 566, A56
  
\bibitem[{{The LIGO Scientific Collaboration} \& {The Virgo Collaboration}(2018){The LIGO Scientific Collaboration} \& {The Virgo Collaboration}}]{LIGO}
{The LIGO Scientific Collaboration} \& {The Virgo Collaboration} 2018, ArXiv
  e-prints, arXiv:1811.12940

\bibitem[{{Lyra} {et~al.}(2010){Lyra}, {Paardekooper}, \& {Mac Low}}]{lyra}
{Lyra}, W., {Paardekooper}, S.-J., \& {Mac Low}, M.-M. 2010, \apjl, 715, L68

\bibitem[{Lyra \& Umurhan(2018)}]{lyra_umurhan}
Lyra, W., \& Umurhan, O. 2018, ArXiv e-prints, arXiv:1808.08681

\bibitem[Masset(2017)]{masset} 
Masset, F.~S., 2017, \mnras, 472, 4204

\bibitem[McNally et al.(2019)]{mcnally2019} 
McNally, C. P., Nelson, R. P., Paardekooper, S.-J., \&
Ben\'{\i}tez-Llambay, P. 2019, \mnras, 484, 728

\bibitem[{{McKernan} {et~al.}(2018){McKernan}, {Ford}, {Bellovary}, {Leigh},
  {Haiman}, {Kocsis}, {Lyra}, {MacLow}, {Metzger}, {O'Dowd}, {Endlich}, \&
  {Rosen}}]{mckernan18}
{McKernan}, B., {Ford}, K.~E.~S., {Bellovary}, J., {et~al.} 2018, ApJ, 866, 66

\bibitem[{{McKernan} {et~al.}(2014){McKernan}, {Ford}, {Kocsis}, {Lyra}, \&
  {Winter}}]{mckernan14}
{McKernan}, B., {Ford}, K.~E.~S., {Kocsis}, B., {Lyra}, W., \& {Winter}, L.~M.
  2014, \mnras, 441, 900

\bibitem[{{McKernan} {et~al.}(2012){McKernan}, {Ford}, {Lyra}, \&
  {Perets}}]{mckernan2012}
{McKernan}, B., {Ford}, K.~E.~S., {Lyra}, W., \& {Perets}, H.~B. 2012, \mnras,
  425, 460
  
\bibitem[{{Moody} {et~al.}(2019){Moody}, {Shi}, \&
  {Stone}}]{moody}
{Moody}, M.~S.~L., {Shi}, J.-M., \& {Stone}, J.~M. 2019, \apj,
  875, 66

\bibitem[{{Mordasini} {et~al.}(2017){Mordasini}, {Marleau}, \&
  {Molli\`{e}re}}]{mordasini}
{Mordasini}, C., {Marleau}, G.-D., \& {Molli{\'e}re}, P. 2017, \aap,
  608, 72

\bibitem[{{Nelson} \& {Papaloizou}(2004)}]{nelson}
{Nelson}, R.~P., \& {Papaloizou}, J.~C.~B. 2004, \mnras, 350, 849

\bibitem[{{Ogihara} {et~al.}(2007){Ogihara}, {Ida}, \& {Morbidelli}}]{ogihara}
{Ogihara}, M., {Ida}, S., \& {Morbidelli}, A. 2007, \icarus, 188, 522

\bibitem[{{O'Leary} {et~al.}(2009){O'Leary}, {Kocsis}, \& {Loeb}}]{OLeary}
{O'Leary}, R.~M., {Kocsis}, B., \& {Loeb}, A. 2009, \mnras, 395, 2127

\bibitem[{{Paardekooper}(2014){Paardekooper}}]{paardekooper2014}
{Paardekooper}, S.-J. 2014, \mnras,
  444, 2031

\bibitem[{{Paardekooper} {et~al.}(2010){Paardekooper}, {Baruteau}, {Crida}, \&
  {Kley}}]{paardekooper2010}
{Paardekooper}, S.-J., {Baruteau}, C., {Crida}, A., \& {Kley}, W. 2010, \mnras,
  401, 1950

\bibitem[{{Paardekooper} {et~al.}(2011){Paardekooper}, {Baruteau}, \& {Kley}}]{paardekooper2011}
{Paardekooper}, S.-J., {Baruteau}, C., \& {Kley}, W. 2011, \mnras,
  410, 293

\bibitem[{{Paardekooper} \& {Mellema}(2006)}]{paardekooper_mellema}
{Paardekooper}, S.-J., \& {Mellema}, G. 2006, \aap, 459, L17

\bibitem[{{Paardekooper} \& {Papaloizou}(2009)}]{paardekooper2009}
Paardekooper, S.-J., \& Papaloizou, J. C. B. 2009, \mnras, 394, 2297

\bibitem[{{Peters}(1964)}]{peters}
{Peters}, P.~C. 1964, Phys.\ Rev., 136, 1224

\bibitem[{{Pierens}(2015)}]{pierens}
{Pierens}, A. 2015, \mnras, 454, 2003

\bibitem[{{Rodriguez} {et~al.}(2016){Rodriguez}, {Chatterjee}, \&
  {Rasio}}]{rodriguez}
{Rodriguez}, C.~L., {Chatterjee}, S., \& {Rasio}, F.~A. 2016, \prd, 93, 084029

\bibitem[{{S{\'a}ndor} {et~al.}(2011){S{\'a}ndor}, {Lyra}, \&
  {Dullemond}}]{sandor}
{S{\'a}ndor}, Z., {Lyra}, W., \& {Dullemond}, C.~P. 2011, \apjl, 728, L9

\bibitem[{Sasaki \& Ebisuzaki(2016)}]{sasaki_ebisuzaki}
Sasaki, T., \& Ebisuzaki, T. 2016, Geoscience Frontiers, 8, 215.

\bibitem[{{Schawinski} {et~al.}(2015){Schawinski}, {Koss}, {Berney}, \&
  {Sartori}}]{schawinski}
{Schawinski}, K., {Koss}, M., {Berney}, S., \& {Sartori}, L.~F. 2015, \mnras,
  451, 2517

\bibitem[{{Shakura} \& {Sunyaev}(1973)}]{shakura_sunyaev}
{Shakura}, N.~I., \& {Sunyaev}, R.~A. 1973,\aap, 24, 337

\bibitem[{Sirko \& Goodman(2003)}]{Sirko:2003aa}
Sirko, E., \& Goodman, J. 2003, Mon.Not.Roy.Astron.Soc., 341, 501.
\newblock \url{https://arxiv.org/abs/astro-ph/0209469}

\bibitem[{{Tanaka} {et~al.}(2002){Tanaka}, {Takeuchi}, \& {Ward}}]{tanaka}
{Tanaka}, H., {Takeuchi}, T., \& {Ward}, W.~R. 2002, \apj, 565, 1257

\bibitem[{{Tanaka} \& {Ward}(2004)}]{tanaka_ward}
{Tanaka}, H., \& {Ward}, W.~R. 2004, \apj, 602, 388

\bibitem[{{Thompson} {et~al.}(2005){Thompson}, {Quataert}, \&
  {Murray}}]{thompson}
{Thompson}, T.~A., {Quataert}, E., \& {Murray}, N. 2005, \apj, 630, 167

\bibitem[{{Uribe} {et~al.}(2015){uribe}, {Bans}, \&
  {K{\"o}nigl}}]{uribe}
{Uribe}, A., {Bans}, A., \& {K{\"o}nigl}, A. 2015, \apj, 802, 54

\bibitem[{{Uribe} {et~al.}(2011){Uribe}, {Klahr}, {Flock}, \&
  {Henning}}]{uribe2011}
{Uribe}, A.~L., {Klahr}, H., {Flock}, M., \& {Henning}, Th. 2011, \apj, 736, 85

\bibitem[{{Ward}(1997)}]{ward}
{Ward}, W.~R. 1997, \icarus, 126, 261

\bibitem[{{Wardle}(1999)}]{wardle}
{Wardle}, M. 1999, \apjl, 525, L101

\end{thebibliography}

\end{document}